\documentclass[traditabstract]{aa}  
\usepackage{graphicx}
\usepackage{txfonts}
\usepackage{natbib}
\usepackage{longtable}

\begin{document}

\title{Galactic arch{ae}ology of a thick disc: Excavating \object{ESO~533-4} with VIMOS \thanks{Based on observations made at the European Southern Observatory using the Very Large Telescope under programme 091.B-0228(A).}} 

   \author{S.~Comer\'on
          \inst{1,2},
          H.~Salo
          \inst{1},
          J.~Janz
          \inst{3},
          E.~Laurikainen
          \inst{1}, and
          P.~Yoachim
          \inst{4}
          }

   \institute{University of Oulu, Astronomy and Space Physics, P.O.~Box 3000, FI-90014, Finland\\
              \email{seb.comeron@gmail.com}
         \and
             Finnish Centre of Astronomy with ESO (FINCA), University of Turku, V\"ais\"al\"antie 20, FI-21500, Piikki\"o, Finland
          \and
          Centre for Astrophysics and Supercomputing, Swinburne University, Hawthorn, VIC 3122, Australia
          \and Department of Astronomy, University of Washington, Box 351580, Seattle, WA 98195, USA
             }

\authorrunning{Comer\'on, S., et al.}
\titlerunning{Galatic arch{ae}ology of the ESO~533-4 thick disc}

\abstract{The disc of galaxies is made of the superposition of a thin and a thick disc. Star formation is hosted in the thin discs and contributes to their growth. Thick discs are formed of old stars. The formation mechanisms of thick discs are under discussion. Thick discs might have formed either at high redshift on a short timescale or might have been built slowly over a Hubble-Lema\^itre time. They may have an internal or an external origin. Here we adopt a galactic arch{ae}ology approach to study the thick disc of ESO~533-4, i.e. we study  the kinematics and the stellar populations of this galaxy in detail. ESO~533-4 is a Southern, nearby, and almost bulgeless galaxy.

We present the first ever Integral Field Unit spectroscopy of an edge-on galaxy with enough depth and quality to study the thick disc. We exposed ESO~533-4 with the blue grism of the VIMOS instrument of the VLT for 6.5\,hours. The field of view covered an axial extent from $\sim0.1\,r_{25}$ to $\sim0.7\,r_{25}$, where $r_{25}$ is the $25\,{\rm mag\,arcsec^{-2}}$ isophotal radius. This corresponds to the range from $\sim1$\,kpc to $\sim7$\,kpc. We used pPXF and the MILES library to obtain velocity and stellar population maps. We compared our kinematic data with simple GADGET-2 models.

The apparent rotational lag of the thick disc of ESO~533-4 is compatible with that expected from the combinations of two effects: differential asymmetric drift and the projection effects arising from studying a disc a few degrees ($2-3^{\rm o}$) away from edge-on. Thus, ESO~533-4 contains little or no counter-rotating material. This is compatible with three formation scenarios: the secular dynamical heating of an initially thin disc, the formation of the thick disc at high redshift in an early turbulent disc phase, and the creation of a thick disc in a major merger event.  If this last mechanism occurred   in all galaxies, it would cause retrograde thick discs in half of them. These retrograde discs have not been observed in the five massive disc galaxies (circular velocity $v_{\rm c}\gtrsim120\,{\rm km\,s^{-1}}$) for which the kinematics of the thick disc is known. The stellar populations map indicates that the populations of the thin and the thick discs of ESO~533-4 are possibly separated in the ${\rm Age}-{\rm log}\,(Z/Z_{\bigodot})$ plane. This would imply that  thin and  thick discs are formed of two distinct stellar populations. The stellar population results are not fully conclusive because of the high dust extinction in ESO~533-4 and because recovering stellar populations is a difficult inverse problem. Having said that, the stellar population results do not favour a secular evolution origin for the thick disc. Hence, we suggest that the thick disc of ESO~533-4 formed in a relatively short event.}


\maketitle

\section{Introduction: Thick discs and galactic arch{ae}ology}

\begin{table*}
 \caption{Summary of the signatures predicted in the thick disc formation scenarios}
 \label{scenarii}
 \centering
 \begin{tabular}{l c c c c}
 \hline\hline
Scenario & External/ & Fast/  & Counter-rotating material & Chemical and/or age discontinuity\\
         & Internal & Secular & in the thick disc       & between the thin and thick discs \\
\hline
Merger of gas-rich galaxies & External & Fast & Possibly\tablefootmark{a} & Yes\\
Turbulent early disc        & Internal & Fast & No\tablefootmark{b}  & Yes\\
Accretion of stars from satellites& External & Secular & Possibly\tablefootmark{a} & Yes\\
Heating by satellites & External & Secular & No & No\\
Internal heating & Internal & Secular & No & No\\
 \hline
 \end{tabular}

 \tablefoottext{a}{Those processes would cause some amount of counter-rotating material provided that some of the mergers are retrograde.} \tablefoottext{b}{In the \citet{BOUR14} simulations, some counter-rotating stars are created in this process. The fraction of retrograde stars decreases as a function of the mass of the galaxy. A high-$z$ galaxy with a $1.4\times10^{10}\,{\mathcal M}_{\bigodot}$ baryonic mass, similar to the mass of ESO~533-4's thick disc, would have 2.2\% of counter-rotating stars (F.~Bournaud, private communication). This small fraction of counter-rotating stars would be undetectable with the techniques used in this paper.}

\end{table*}

We know with absolute certainty that our universe contains galaxies \citep{CUR17, OP22, HUB25}. Another fact established with almost the same degree of certitude is that the initial state of the universe was a hot Big Bang \citep{LE31, GA46, ALPH49, PEN65}. It follows that the products of the primordial nucleosynthesis somehow assembled and evolved into the galaxies that we observe roughly a Hubble-Lema\^itre time after creation. The current cosmological paradigm is Lambda cold dark matter \citep[$\Lambda$CDM; ][for a recent parameter determination]{A15}, where the energy budget of the universe is dominated by elusive dark matter and a positive cosmological constant. Within that paradigm, the evolution of galaxies is driven by a combination of interactions with the environment \citep[nurture; e.g.][]{TOOM77} and internal evolution \citep[nature; e.g.][]{KOR04, ATH13}. However, the details and the specific weight of each of these two families of processes in the galaxy construction and subsequent evolution still elude us.

One of the most mysterious remnants of galaxy formation and evolution are thick discs. Seen as roughly exponential vertical excesses of light in edge-on galaxies a few thin disc scale heights above the galaxy mid-planes, their origin is still a matter of debate more than three decennia after their discovery \citep{BURS79, TSI79}. Nowadays, all disc galaxies where thick discs have been searched for have been found to host one or even two of them \citep[][but see \citet{STREICH15}]{YOA06, CO11A, CO11B}. The following scenarios have been proposed to explain their formation:

\begin{itemize}
 \item {{\bf External fast processes:} the thick disc is the consequence of the merger of two or more gas-rich galaxies during the initial assembly process \citep{BROOK04}.}
 \item {{\bf Internal fast processes:} the thick disc is born thick because of the turbulent and clumpy nature of the first discs \citep{EL06, BOUR09, CO14}.}
 \item {{\bf External secular processes:} the thick disc is made of stars stripped from infalling satellites \citep{AB03} and/or by the disc dynamical heating caused by these satellites \citep{QUINN93, QU11}. We call these processes secular because minor mergers are much slower than major mergers as a result of  weaker dynamical friction. Also, according to $\Lambda$CDM, several of these events can occur in a Hubble-Lema\^itre time \citep[e.g. several tens of minor mergers causing tidal features are expected in][]{JOHN08}.}
 \item {{\bf Internal secular processes:} the thick disc is caused by dynamical heating due to disc overdensities \citep{VILL85} and/or the radial migration of stars \citep{SCHO09A, SCHO09B, LOEB11}. Radial migration as a viable mechanism to build thick discs is contested by \citet{MIN12} and \citet{VE14}.}
\end{itemize}

It is likely that a combination of two or more of these scenarios contributed to the thick disc formation. Actually, some mechanisms, such as secular internal heating, must always occur to some extent. In \citet{CO12}, we suggested that thick discs in low- and high-mass galaxies might have different dominant formation mechanisms. This is because there is a clear bimodality in the ratio of thick to thin disc masses as a function of the total galaxy mass: high-mass galaxies have the same ratio irrespective of the galaxy total mass, whereas for low-mass galaxies the ratio decreases as the total galaxy mass increases. The fact that most high-mass galaxies have roughly similar thick to thin disc mass ratios \citep{CO14} suggests the possibility that the same combination of mechanisms  built them. The dividing line between low- and high-mass galaxies is found at a circular velocity $v_{\rm c}\approx120\,{\rm km\,s^{-1}}$ \citep[suggested by][]{YOA06, YOA08A}, which corresponds to a baryonic mass of $\mathcal{M}\sim10^{10}\,\mathcal{M}_{\bigodot}$ when using the Tully-Fisher relation in \citet{ZAR14}.

A particular model \citep{SCHO09A, SCHO09B, ROS13, MIN15}, which proposes a thick disc created through a combination of internal and external secular heating mechanisms, is relevant in later sections of the paper. In this model, each of the mono-age populations of a galaxy is distributed in a flared disc. The younger stellar populations have a longer scale length than the older stellar populations. The superposition of a series of flares at different radii produce what has been identified in photometric decompositions
as a thick disc. The result of this model is a galaxy with no apparent flare even if the individual mono-age populations do flare.

As a result of cosmological dimming and resolution problems, it is often impractical to tackle galaxy evolution issues by observing highly redshifted objects. An alternative is so-called galactic arch{ae}ology, which consists of studying the structures of redshift $z\sim0$ galaxies in great detail to infer their origin. Thick discs are ideal targets for galactic archaeology because different formation scenarios predict distinct kinematical and chemical signatures, as summarized in Table~\ref{scenarii}.

So far, attempts to do galactic arch{ae}ology in external edge-on galaxies to study thick discs have been mostly based on photometric decompositions \citep{YOA06, CO11C, CO12}. The exception is the work by \citet{YOA08A,YOA08B}, in which single-slit spectroscopy was used to study  thin and  thick discs. They found that  thick discs are made of old stars and that  thick discs of some low-mass galaxies contain counter-rotating stars. None of the three massive galaxies in their sample had measurable amounts of retrograde stars. The problem that arises from single-slit observations is that one has to rely on photometric decompositions to find the range of heights where the thick disc dominates before placing the slit. Ideally, we would like to be able to observe galaxies in a wide range of heights, so no assumptions about the extent of the thick disc have to be made a priori.

\section{ESO~533-4}

\begin{table}
 \caption{Parameters of ESO~533-4}
 \label{summary}
 \centering
 \begin{tabular}{l c c}
 \hline\hline
 Parameter & Value & Source\\
 \hline
 RA (J2000.0) &$22^{\rm h}14^{\rm m}03\fs02$&1\\
 Dec (J2000.0)& $-26\degr56\arcmin15\farcs8$&1\\
 $M_{3.6\,\mu{\rm m}}({\rm AB})$ & $-20.52$ & 2\\
 Type         & Sc? sp                      &3 \\
              & S0$^{\rm o}_{\rm c}$ sp/E(d)8   & 4\\
 $v_{\rm sys}$ (helicoentric) & $2596\pm3\,{\rm km\,s^{-1}}$ & 5\\
 $D$          & $39.8\,{\rm Mpc}$ & 6\\
 $1\arcsec$ physical size& 193\,pc&6\\
 $v_{\rm c}$   & $147.5\pm2.2\,{\rm km\,s^{-1}}$ & 7\\
 $r_{25}$   & $54\farcs6$ & 7\\
 Luminosity profile & Type~2 break & 8\\
 $\mathcal{M}_{\rm g}$ & $(680\pm20)\times10^7\,\mathcal{M}_{\bigodot}$ & 9\\
 $\mathcal{M}_{\rm CMC}$ & $(19\pm2)\times10^7\,\mathcal{M}_{\bigodot}$ & 9\\
 $\mathcal{M}_{\rm t}$ & $(1500\pm300)\times10^7\,\mathcal{M}_{\bigodot}$ & 10\\
 $\mathcal{M}_{\rm T}$ & $(1100\pm200)\times10^7\,\mathcal{M}_{\bigodot}$ & 10\\
 \hline
 
 \end{tabular}
 \tablebib{(1) \citet{SKRU06}; (2) \citet{MU15}; (3) \citet{VAU91}; (4) \citet{BU15}; (5) NED; (6) from NED average of redshift-independent distances; (7) HyperLeda \citep{MAK14}; (8) \citet{CO12}; (9) atomic gas disc mass ($\mathcal{M}_{\rm g}$) and CMC mass ($\mathcal{M}_{\rm CMC}$) from \citet{CO14}; (10) thin disc mass ($\mathcal{M}_{\rm t}$) and thick disc mass ($\mathcal{M}_{\rm T}$) from this paper.}
\end{table}

\begin{figure}
\begin{center}
  \includegraphics[width=0.48\textwidth]{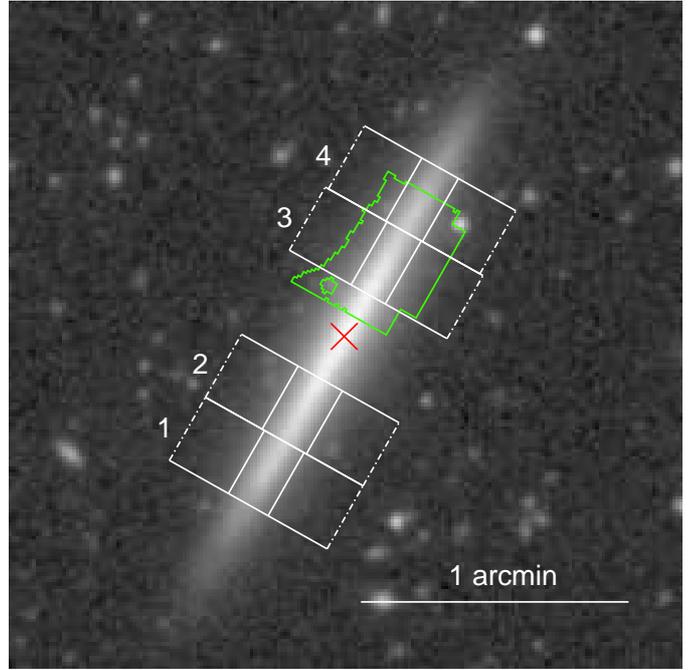}
  \end{center}
  \caption{\label{ESO533-4} $3.6\,\mu{\rm m}$ image of ESO~533-4 from the S$^4$G. The red cross indicates the centre of the galaxy and the region with a green border is that we studied. North is up and east is left. The white solid lines perpendicular to the galaxy mid-plane show the four axial bins used for the luminosity profile fits in Fig.~\ref{thinthick}. The dash-dotted lines indicate the heights over which the profiles were fit. The solid lines parallel to the galaxy mid-plane correspond to the height above which 90\% of the light comes from the thick disc. The numbers are used to associate the bins with the fits in Fig.~\ref{thinthick}.}  
\end{figure}

We present the first integral field spectroscopy study of the thin/thick structure of an edge-on disc. We used the VIsible Multi-Object Spectrograph of the Very Large Telescope \citep[VIMOS;][]{LE03}. The selected target was ESO~533-4. This galaxy has been part of the sample of our recent thick disc studies \citep{CO12, CO14}. The target was chosen because of:
\begin{itemize}
\item its small central mass concentration (CMC), and hence, presumably little effect of a central spheroid on the thin and thick disc kinematics;
\item its relatively small angular size, so a large fraction of the disc could be covered by the VIMOS field of view (FOV); and\item a brighter than average thick disc, which  starts to dominate at $m_{3.6\,\mu{\rm m}}({\rm AB})\approx22\,{\rm mag\,arcsec^{-2}}$ \citep{CO12} so that thick disc spectra can be obtained with a few hours of exposure.
\end{itemize}

A summary of the properties of ESO~533-4 is presented in Table~\ref{summary}. There is a disagreement about the stage of the galaxy in the literature. ESO~533-4 has a small CMC, which  makes it appear late type \citep[Sc;][]{VAU91}, but its smooth appearance at 3.6\,$\mu{\rm m}$ has recently caused it to be classified as an S0$^{\rm o}_{\rm c}$ \citep{BU15}. ESO~533-4 has a baryonic mass $\mathcal{M}\sim3\times10^{10}\,\mathcal{M}_{\bigodot}$, and hence it is a high-mass galaxy. A $3.6\,\mu{\rm m}$ image from the Spitzer Survey of Stellar Structure in Galaxies \citep[S$^4$G;][]{SHETH10} is presented in Fig.~\ref{ESO533-4}.

\section{The thin/thick disc decomposition of ESO~533-4}

\begin{figure*}
\begin{center}
  \begin{tabular}{c c}
  {\large Bin~1}&{\large Bin~2}\\
  {\large $-0.8\,r_{25}<R<-0.5\,r_{25}$}&{\large $-0.5\,r_{25}<R<-0.2\,r_{25}$}  \vspace{-1.5mm}\\
  \includegraphics[width=0.48\textwidth]{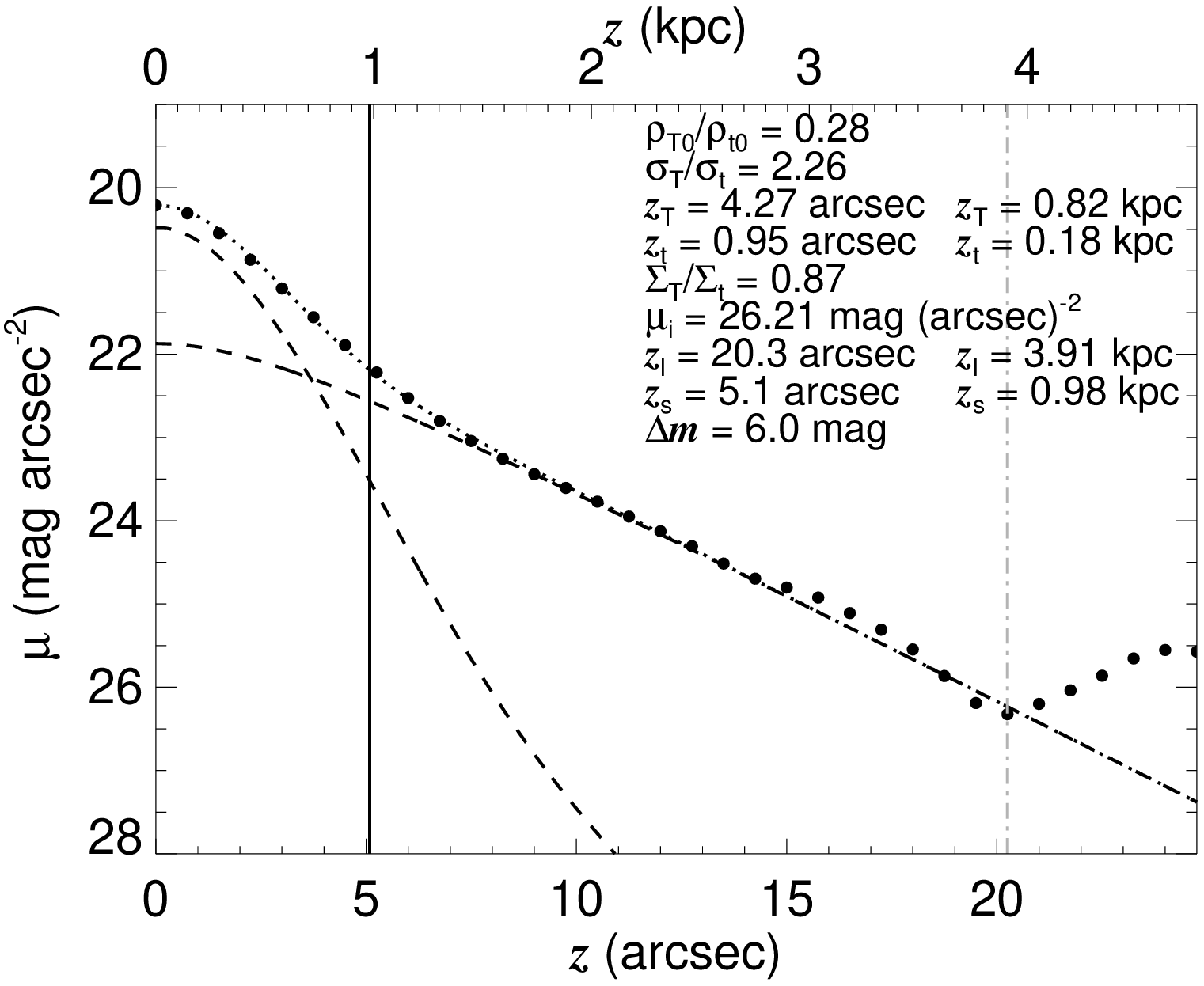}&
  \includegraphics[width=0.48\textwidth]{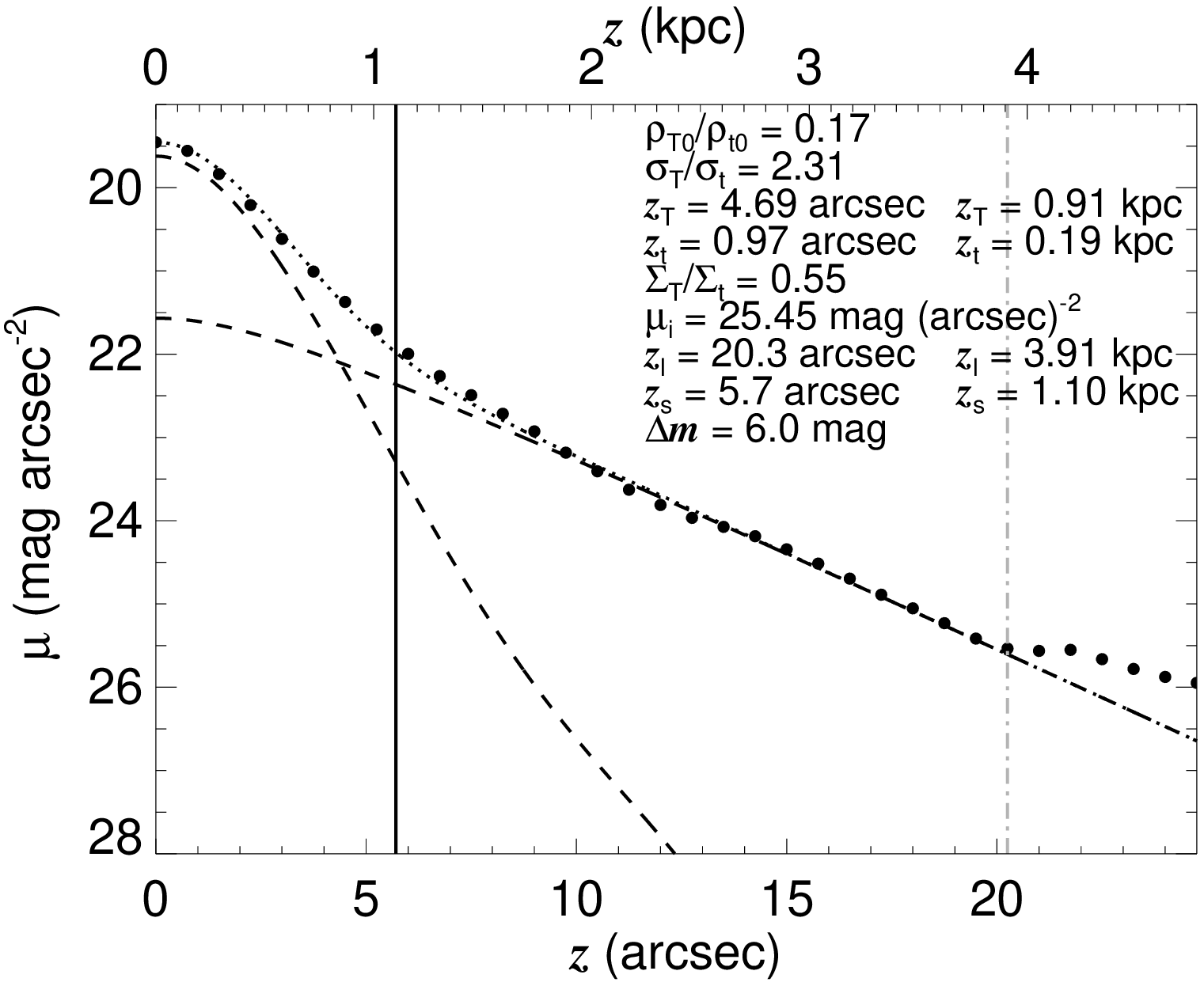} \vspace{4mm}\\
  {\large Bin~3}&{\large Bin~4}\\
  {\large $0.2\,r_{25}<R<0.5\,r_{25}$}&{\large $0.5\,r_{25}<R<0.8\,r_{25}$}\vspace{-1.5mm}\\
  \includegraphics[width=0.48\textwidth]{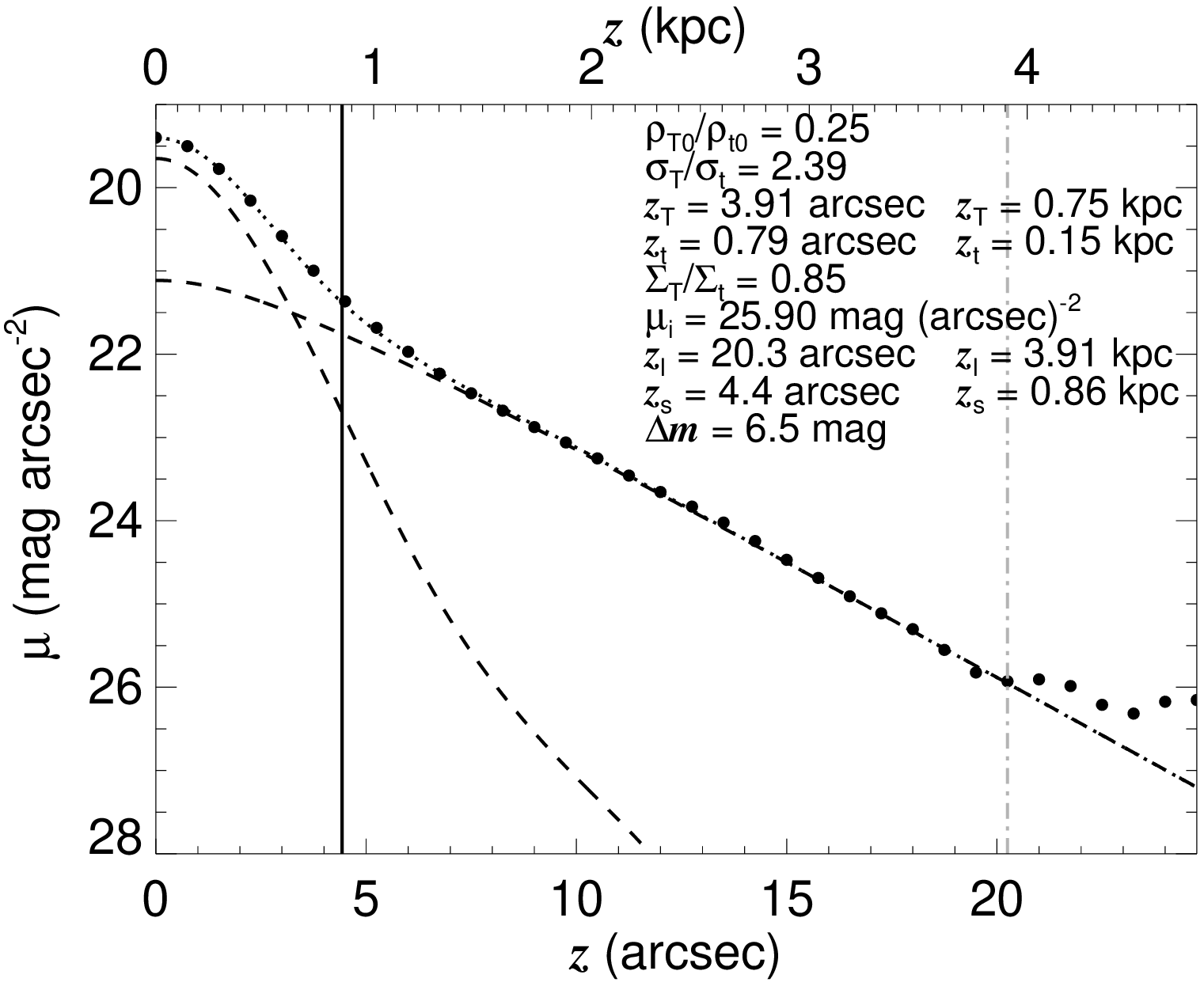}&
  \includegraphics[width=0.48\textwidth]{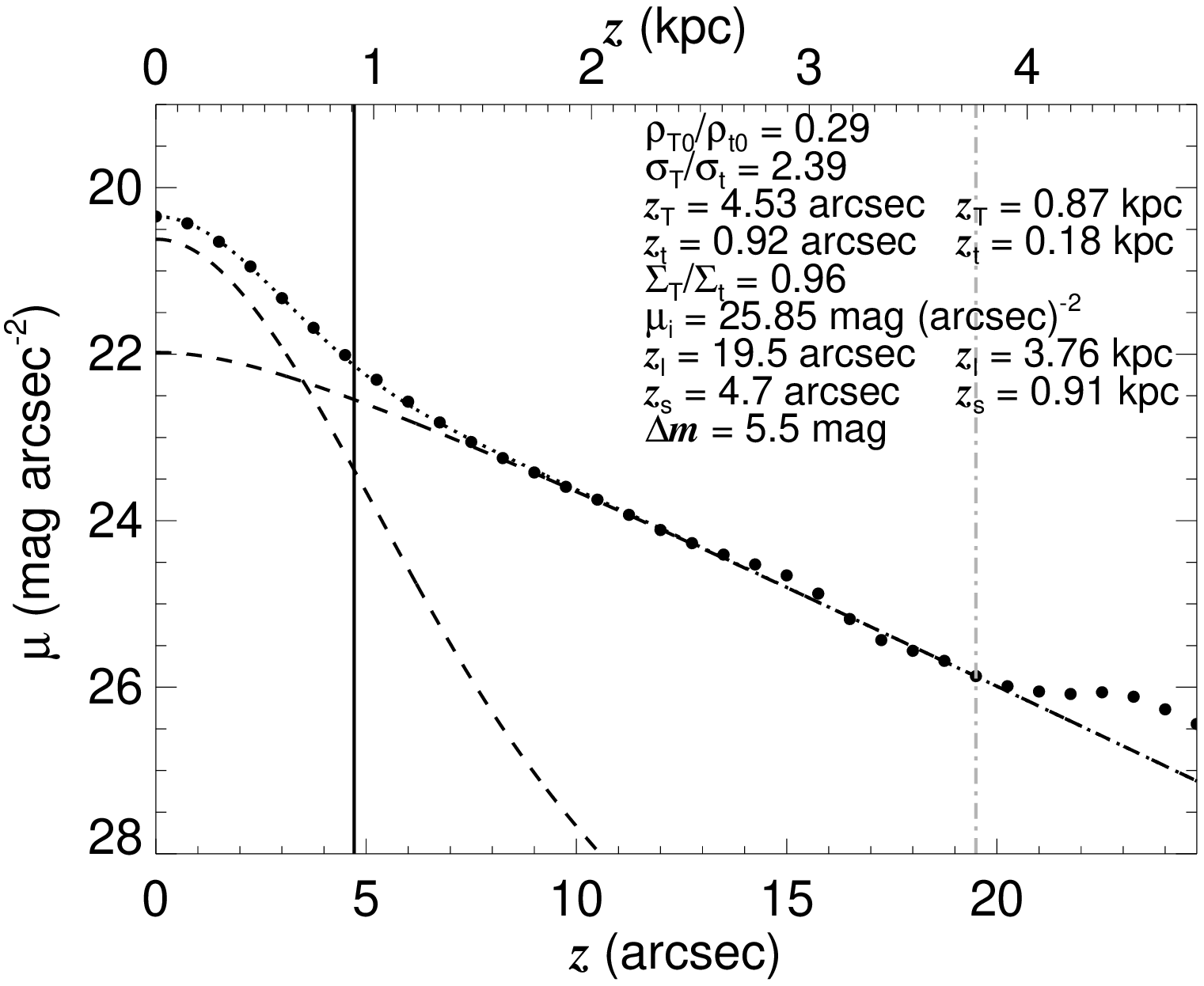}\\
  \end{tabular}
  \end{center}
  \caption{\label{thinthick} Luminosity profiles of ESO~533-4 in the four bins perpendicular to the mid-plane that are indicated in Fig.~\ref{ESO533-4} (large dots). The fits are indicated with the dotted lines, and the contribution of the thin and  thick discs are indicated with dashed lines. The vertical solid line indicates the height, $z_{\rm s}$, above which 90\% of the light comes from the thick disc according to our fit. The dot-dashed grey vertical line indicates the maximum height, $z_{\rm l}$, that we use for the fit. $\rho_{\rm T0}/\rho_{\rm t0}$ stands for the thick to thin disc mass density ratio at the mid-plane, $\sigma_{\rm T}/\sigma_{\rm t}$ for their velocity dispersion ratio, $z_{\rm T}$ and $\rm z_{\rm t}$ for the thick and thin disc vertical scale heights, $\Sigma_{\rm T}/\Sigma_{\rm t}$ for the thick to thin disc mass ratio, $\mu_{\rm i}$ for the lowest fitted surface brightness level, and $\Delta m$ for the range in magnitudes over which the fit was done.}  
\end{figure*}

In \citet{CO12}, we made thin/thick disc decompositions of ESO~533-4 for the symmetrized $3.6\,\mu{\rm m}+4.5\,\mu{\rm m}$ luminosity profiles in four bins perpendicular to the galaxy mid-plane. The discs were assumed to be in hydrostatic equilibrium following the formalism by \citet{NA02}. We accounted for the gravitational effect of a thin gas disc with 20\% of the surface mass density of the thin disc. We made several simplifying assumptions,  such as similar scale lengths for the thin and  thick discs, that the disc is not  too submaximal in its inner parts, vertical isothermality, and roughly constant scale heights with changing radii.

In \citet{CO12} we convolved our solutions with a Gaussian kernel to account for the point spread function (PSF) of the S$^4$G images. The kernel had a $2\farcs2$ full width at half maximum (FWHM). However, the S$^4$G PSF is now known to have faint but extended wings not accounted for by a Gaussian. Here we redo the fits  using the S$^4$G symmetrized PSF \citep{SA15}. Here we produce our fit using only the $3.6\,\mu{\rm m}$ band because we only have a detailed PSF model for that band.

We use the sky values from the S$^4$G Pipeline~3 \citep{MU15} instead of our own sky determinations. To do the fit, we use {\sc idl}'s {\sc curvefit} instead of our own minimization routine \citep[described in][]{CO11C}. As in \citet{CO12}, we use a ratio of mass-to-light ratios of the thick and thin discs $\Upsilon_{\rm T}/\Upsilon_{\rm t}=1.2$. The $\Upsilon_{\rm T}/\Upsilon_{\rm t}=1.2$ value comes from studying the spectral energy distributions resulting from the Milky Way star formation history models by \citet{NY06} for the thin and the thick discs \citep{CO11C}. The maximum height of the fit was selected as that beyond which it is not possible to make a two disc component fit (the mean squared difference between the observed profile and the fit becomes larger than $0.01\,({\rm mag\,arcsec^{-2}})^2$ at that height). What limits the height of the fitting region is the presence of scattered light of point sources in the field and background gradient issues. The extent of the fitting range covers over four thick disc scale heights.

The new thin/thick disc fits are very similar to those in \cite{CO12}, which indicates that the errors introduced by approximating the PSF with a Gaussian function were small. The symmetrized PSF used here covers $15^{\prime\prime}$ in radius. The thick disc dominates the luminosity profiles down to $z\approx5^{\prime\prime}$ ($\sim1$\,kpc), which is a height safely smaller than the radius of the modelled PSF. If we consider our fitted thin disc, we find that the extra light added to $|z|>5^{\prime\prime}$ is about $4\%$ of the observed emission. Therefore, we can discard the possibility that the thick disc is an artefact caused by scattered light as suggested by \citet{SAN15}.

The new fits indicate a slightly higher ratio of surface densities between the thick and  thin disc ($\Sigma_{\rm T}/\Sigma_{\rm t}$) compared to \citet{CO12}, which slightly increases the mass ratio of the two discs: $\mathcal{M}_{\rm T}/\mathcal{M}_{\rm t}=0.65$ in \citet{CO12} and $\mathcal{M}_{\rm T}/\mathcal{M}_{\rm t}=0.77$ here, following Eq.~5 in \citet{CO11C}. We could use a gas disc surface  density larger than 20\% of that of the thin disc to reflect the fact that the gas mass is $\sim45\%$ that of the thin disc (Table~\ref{summary}). However, this would not take into account that usually gas discs are more extended than stellar discs. Furthermore, when we refitted the luminosity profiles with double the gas surface density, we obtained the same $\mathcal{M}_{\rm T}/\mathcal{M}_{\rm t}$ within 1\%.

We find that 90\% of the light coming from the region $|z|\gtrsim5^{\prime\prime}$ ($z\gtrsim970$\,pc) is emitted by the thick disc. In what follows, we consider that region as that where the thick disc dominates the surface brightness.
From our fits, we obtain that 39\% of the light comes from the thick disc. In \citet{SA15}, this fraction is 33\% when using {\sc Galfit3.0} \citep{PENG10} for the decomposition ({\sc Galfit3.0}, though performing a simultaneous 2D fit over the whole galaxy, oversimplifies the luminosity profiles perpendicular to the mid-planes by assuming that the discs have vertical ${\rm sech}^2(z/z_0)$ profiles). The {\sc Galfit3.0} thick disc scale height is slightly underestimated ($z_{\rm T}=3\farcs4$ for {\sc Galfit3.0} versus $z_{\rm T}=4\farcs4$ in our determination when averaging over the four fits in Fig.~\ref{thinthick}). {\sc Galfit3.0} finds that the height above which 90\% of the light comes from the thick disc is $z=4\farcs6$.

\section{Observations}

The spectroscopic observations were made on the nights $6-7$ and $7-8$ of August 2013. We used the integral field unit mode of VIMOS with the High Resolution Blue grism and a spacial resolution of $0\farcs66\times0\farcs66$ per fibre. The FOV of this mode is $27\arcsec\times27\arcsec$. The wavelength coverage of this configuration is $3700-5350\,\AA$ and the dispersion is $0.71\,\AA\,{\rm pixel}^{-1}$.

The total in-target exposure was 6.5\,hours (2\,hours in the first night and 4.5\,hours during the second night). The exposure was divided into 13 observing blocks (OBs) with a few arcsecond dithering between them. Each OB exposed ESO~533-4 for 30\,minutes. We also took 4\,minute exposures of the sky, at a $2\arcmin$ distance in the direction perpendicular to the mid-plane. During the first night, a sky exposure was taken for every two galaxy OBs. During the second night two sky exposures were taken for every three galaxy OBs. Both nights were photometric and dark. The seeing was large ($\gtrsim1\arcsec$), but this does not affect our science because in any case we need a large spacial binning to study the thick disc with a sufficiently large signal-to-noise ratio ($S/N$).
Standard calibration exposures (flat fields, bias, calibration lamps) and spectrophotometric standard stars were taken for the two nights.

\section{Data reduction}

\subsection{Basic reduction}

First, we used {\sc idl}'s {\sc la\_cosmic} routine to clean the cosmic rays.
The spectra were reduced and flux calibrated by the VIMOS pipeline. The Reflex environment \citep{FREUD13} was used to run the pipeline. VIMOS has four quadrants and the pipeline provides independent reductions for each quadrant. Therefore, the output of the initial reduction was $4\times13=52$ sets of spectra corresponding to the galaxy.

For each of the quadrants of the sky exposures, we created a single spectrum by averaging the spectra in each spaxel with a $3\sigma$ clipping. We calculated the sky emission for each of our galaxy exposures by associating them with the average of the sky spectra from the same quadrant taken within 64\,minutes. This time was set so at least one sky exposure could be linked to each galaxy exposure, but in some cases there were up to four. Each of the quadrants has a remarkably uniform sky background. Quadrant number~2 has $\sim40\%$ less signal than the others.

The information in headers could not be used for spacial alignment because of a few arcsecond imprecisions in the pointing. To circumvent this problem, we tried to align the frames by finding point sources in images made by collapsing the spectra. However, the two point sources in the vicinity of the pointings were close to the edge of the field, and thus did not show up in some of the exposures as a result of the ditherings. Then, we built [O{\sc iii}] images by integrating in the spectral direction from $\lambda=5047.5\,\AA$ to $\lambda=5054.6\,\AA$ (this takes  the redshift of ESO~533-4 into account). Unresolved star-forming regions stand out in the [O{\sc iii}] images and were used for manual alignment.
Once the galaxy frames were spacially aligned, we summed the spectra for every spaxel using a $3\sigma$ clipping. We also summed for each spaxel the sky spectra associated with the galaxy frames that were not clipped away. For every wavelength in each spectrum corresponding to each spaxel, we estimated the $S/N$ by considering the Poisson noise and the readout noise,
\begin{equation}
S/N=\frac{S_{\rm galaxy}}{\sqrt{S_{\rm galaxy}+S_{\rm sky}+\sum_{i}{\left({RON_{i}}^2\right)}}} 
\end{equation}
where, $S_{\rm galaxy}$ is the signal from the galaxy, $S_{\rm sky}$ is the signal from the sky, and $RON_{i}$ are the read-out noises of each of the exposures that were summed to obtain the final spectrum of a given spaxel. We neglected the dark current contribution to the $S/N$ because it is very small (typically $5\,e^{-}\,{\rm hr^{-1}\,pix^{-1}}$, according to the VIMOS Pipeline User Manual\footnote{ftp://ftp.eso.org/pub/dfs/pipelines/vimos/vimos-pipeline-manual-6.8.pdf}).

We masked the two point sources in the field, thought to be foreground stars or ESO~533-4 globular clusters, as well as several spaxels in the edges that were covered by few exposures. The two point sources in the field were used to align the IFU data with the S$^4$G data. The alignment precision is estimated to be of the order of $1^{\prime\prime}$.

\subsection{Kinematics}

\subsubsection{Velocity maps}

\begin{figure*}
\begin{center}
  \includegraphics[width=0.95\textwidth]{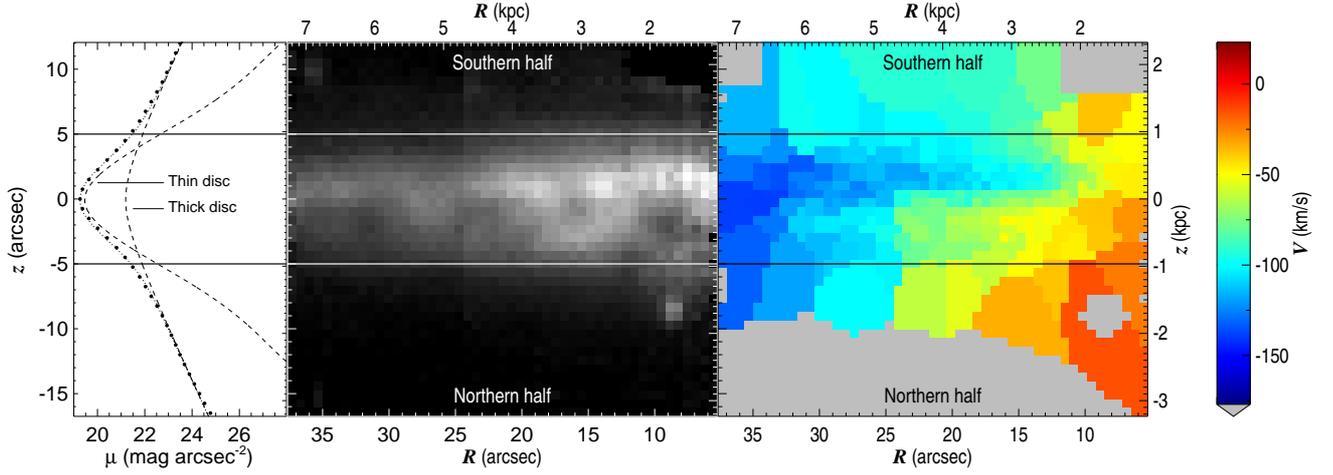}
  \end{center}
  \caption{\label{velocity} Left panel: symmetrized, mid-infrared S$^4$G luminosity profile and thin/thick disc decomposition of the $0.08\,r_{25}-0.69\,r_{25}$ ($0.8-7.3$\,kpc) axial range of ESO~533-4 (the one covered by the IFU). The horizontal solid lines indicate the height, $z_{\rm s}$ above which 90\% of the light comes from the thick disc. Middle panel: image of the observed field made by collapsing the spectra. Right: velocity field of ESO~533-4 after the subtraction of the recession velocity. The horizontal axis indicates the axial distance to the galaxy centre and the vertical axis indicates the distance from the mid-plane.}  
\end{figure*}

Spacial binning is necessary to achieve a $S/N$ good enough to study thick disc kinematics. The binning was done with the Voronoi binning code by \citet{CAP03} with the condition of $S/N\sim25$ for spectral pixels between $\lambda=4500\,\AA$ and $\lambda=5000\,\AA$.

To find the velocity field of ESO~533-4 we fitted the spacially binned spectra with the MILES stellar population synthesis models \citep{VAZ10} using the penalized pixel-fitting (pPXF) code by \citet{CAP04}\footnote{Both pPXF and the Voronoi tesselation codes can be found at http://www-astro.physics.ox.ac.uk/$\sim$mxc/software/}. It is usually required to Gauss-convolve either the library or the observed spectra so their FWHMs match. However, we found that the FWHM in our spectra is $\Delta\lambda\approx2.2\,\AA$ on the blue side and $\Delta\lambda\approx2.8\,\AA$ on the red side, which is very similar to MILES' FWHM, $\Delta\lambda=(2.51\pm0.07)\,\AA$. We therefore decided not to make a convolution. Because of the relatively low $S/N$ of the bins, we only fitted two momenta. Emission lines were masked with a mask $\Delta v=1000\,{\rm km\,s^{-1}}$ in width. We used an eight degree additive Legendre polynomial to correct the continuum shape for dust reddening and possible photometric calibration problems. The velocity map is shown in the right panel in Fig.~\ref{velocity}. The height above which 90\% of the light comes from the thick disc is $|z|\approx5^{\prime\prime}$. Therefore, the uppermost and the lowermost rows of bins should be dominated by the thick disc light.

\subsubsection{Rotation curves}

\begin{figure}
\begin{center}
  \includegraphics[width=0.48\textwidth]{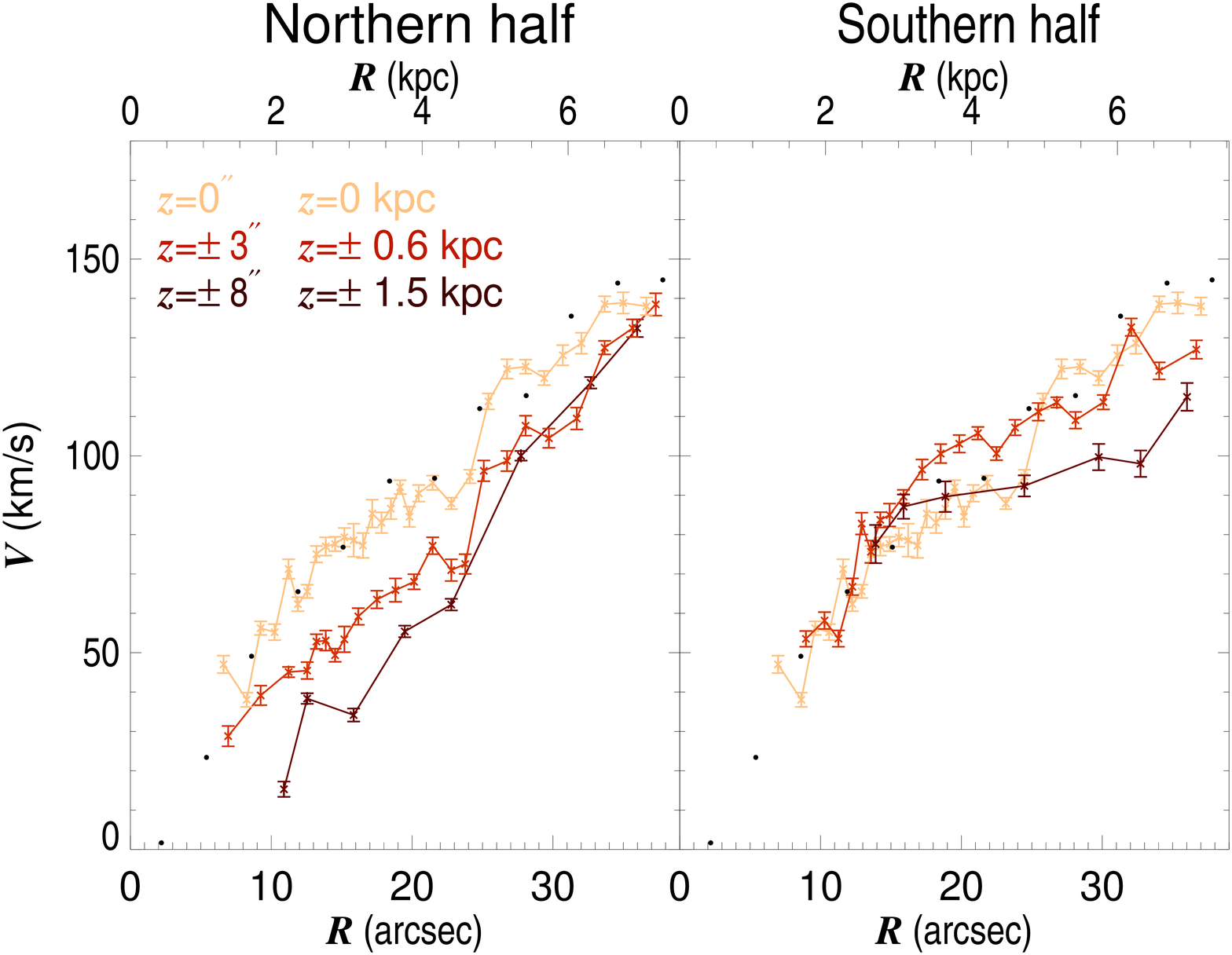}
  \end{center}
  \caption{\label{velprof} Rotation curves of ESO~533-4 at different heights as indicated by the colours. The northern half of the disc corresponds to the bottom half of the disc in Fig.~\ref{velocity}. The black dots correspond to the \citet{MATH92} H$\alpha$ rotation curve with a $+10\,{\rm km\,s^{-1}}$ offset (see text).}  
\end{figure}

We made rotation curves of ESO~533-4 at five different heights, namely $z=0$, $z=\pm3^{\prime\prime}$ ($\sim0.6$\,kpc), and $z=\pm8^{\prime\prime}$ ($\sim1.5$\,kpc; Figure~\ref{velprof}). The $z=\pm8^{\prime\prime}$ cuts should be well within the region where the light emission comes mostly from the thick disc. The profiles take into account the movement of the Earth away from the galaxy ($\sim19\,{\rm km\,s^{-1}}$ at the time of the year at which the observations were performed) and assume a heliocentric systemic velocity $v_{\rm sys}=2596\,{\rm km\,s^{-1}}$ (Table~\ref{summary}). The points in the rotation curves are calculated by making axial cuts in the data used for Fig.~\ref{velocity}. The error bars in the rotation curves are estimated by producing 20 Monte-Carlo simulations (with the estimated noise for each pixel) when obtaining the kinematics from each spectrum.

When extrapolated to $R=0$, the rotation curves are close to $V=0$, which confirms that the spectra calibration process is robust. It is also reassuring that our $z=0$ rotation curve matches very well  the H$\alpha$ rotation curve from \citet{MATH92}, once this has been offset by $+10\,{\rm km\,s^{-1}}$. The correction seems necessary because taken as it is published, the velocity at $R=0$ is about $-10\,{\rm km\,s^{-1}}$. If the offset is not made, the maximum in the H$\alpha$ rotation curve is found at $v\sim160\,{\rm km\,s^{-1}}$ on the western side of the galaxy and at $v\sim140\,{\rm km\,s^{-1}}$ on the eastern side, which would be strange in an undisturbed canonical rotation curve. We suspect an inaccurate calibration or heliocentric correction to be the source of the problems with the H$\alpha$ data as published.

\subsection{Stellar populations}

\begin{figure*}
\begin{center}
  \includegraphics[width=0.95\textwidth]{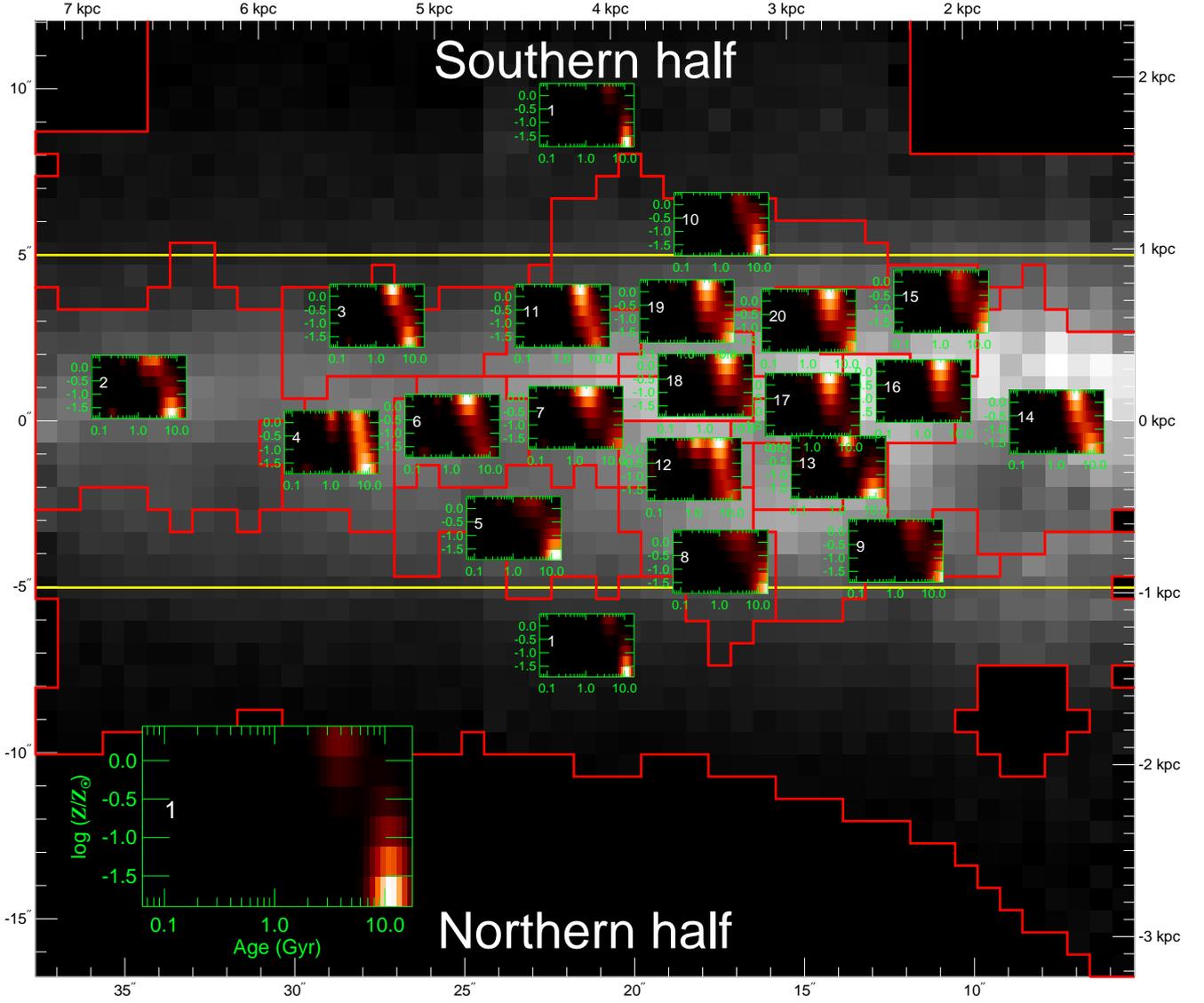}
  \end{center}
\caption{\label{populations} Stellar populations of ESO~533-4. The background image is the same as in the middle panel of Fig.~\ref{velocity}. The red lines indicate the spacial bins used for the analysis. The horizontal solid yellow lines indicate the height above which 90\% of the light comes from the thick disc. On top of each bin there is a stellar population plot with the horizontal axis corresponding to the ages in Gyr and the vertical axis corresponding to the metallicities in ${\rm log}\,\left(Z/Z_{\bigodot}\right)$. The two thick disc sections (uppermost and lowermost tiles) were actually treated as a single bin. The colours in the plots for each bin indicate the mass fraction of a given stellar population. The bottom left corner of the figure shows an enlarged version of one of the stellar population distributions to improve the readability of the axes.}  
\end{figure*}

\begin{figure}
\begin{center}
{Spectrum from the region dominated by the thick disc light}  \vspace{-1.mm}\\
  \includegraphics[width=0.48\textwidth]{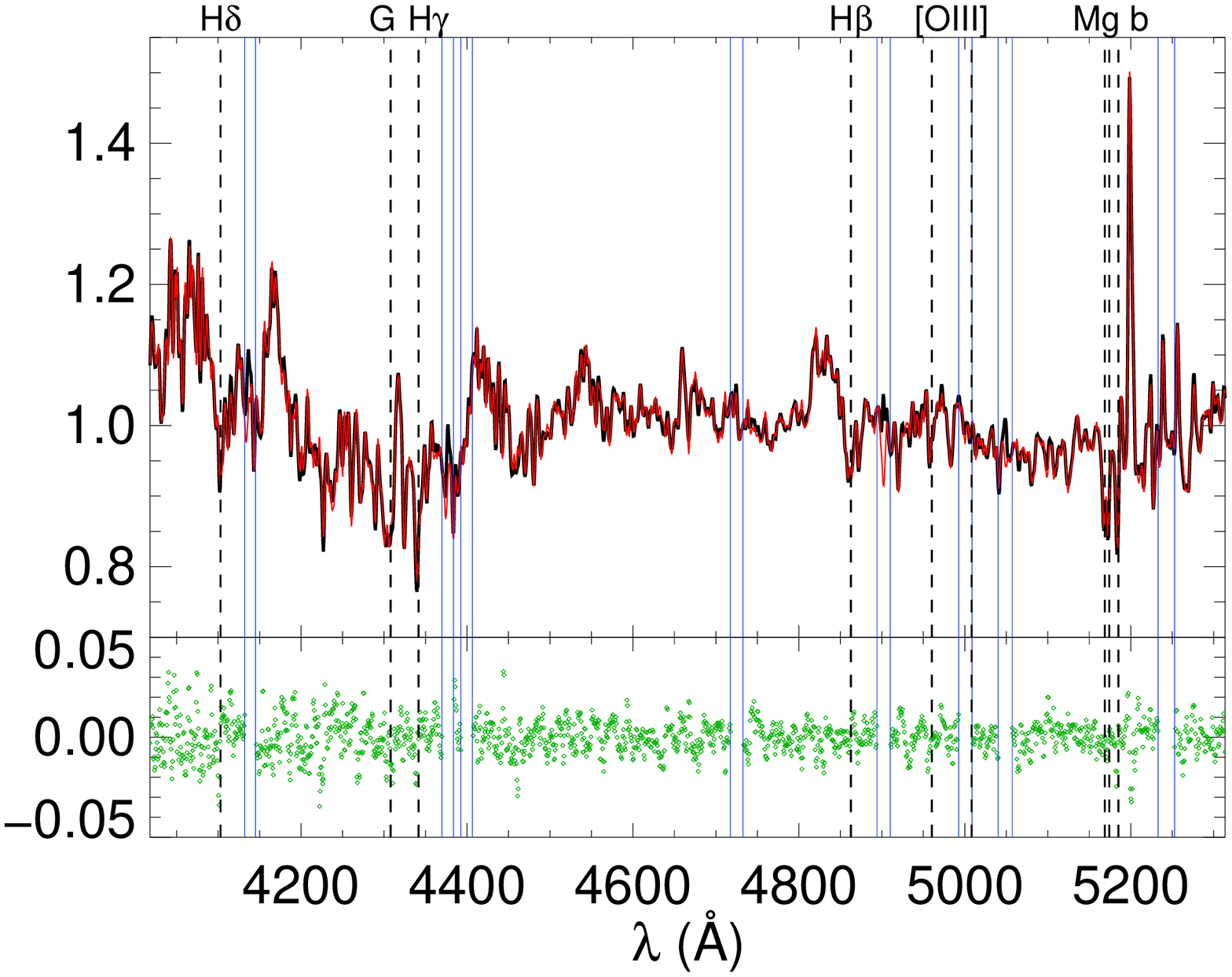}\\
{Spectrum from a region dominated by the thin disc light}  \vspace{-1.mm}\\
  \includegraphics[width=0.48\textwidth]{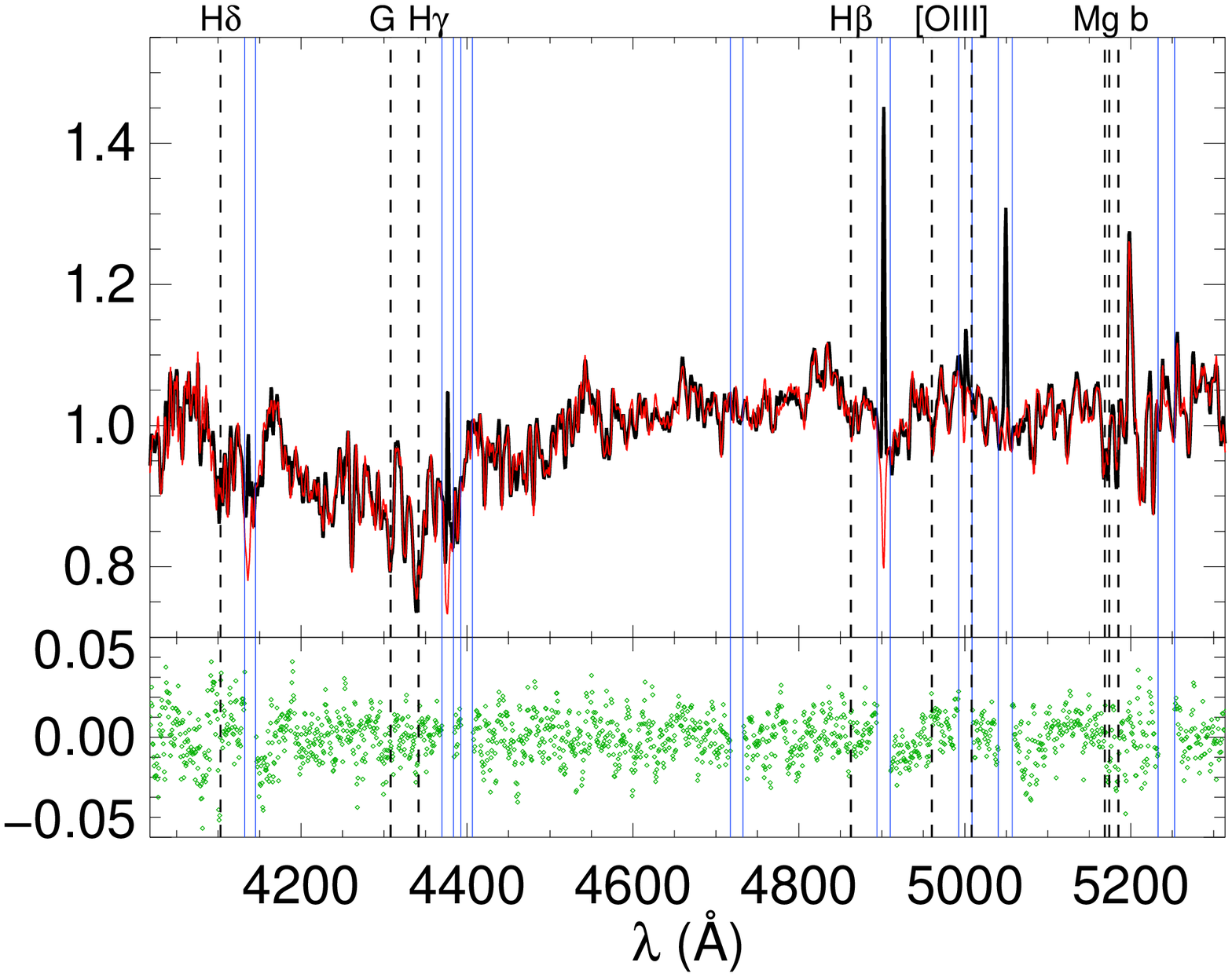}\\
  \end{center}
\caption{\label{examplespec} Two examples of rest-frame spectra obtained from our $S/N\sim80$ tesselation. The black line corresponds to the actual spectrum, the red line corresponds to the fit, and the blue vertical lines indicate the spectral windows that were masked because of the possible presence of emission lines. Residuals are shown with green symbols below the spectra. The top spectrum belongs to the thick disc dominated region and the bottom spectrum to the rightmost thin disc dominated tile in Fig.~\ref{populations}. The spectra are normalized to their median value. Several spectral lines are indicated with dashed lines.}  
\end{figure}

\begin{figure*}
\begin{center}
  \includegraphics[width=0.95\textwidth]{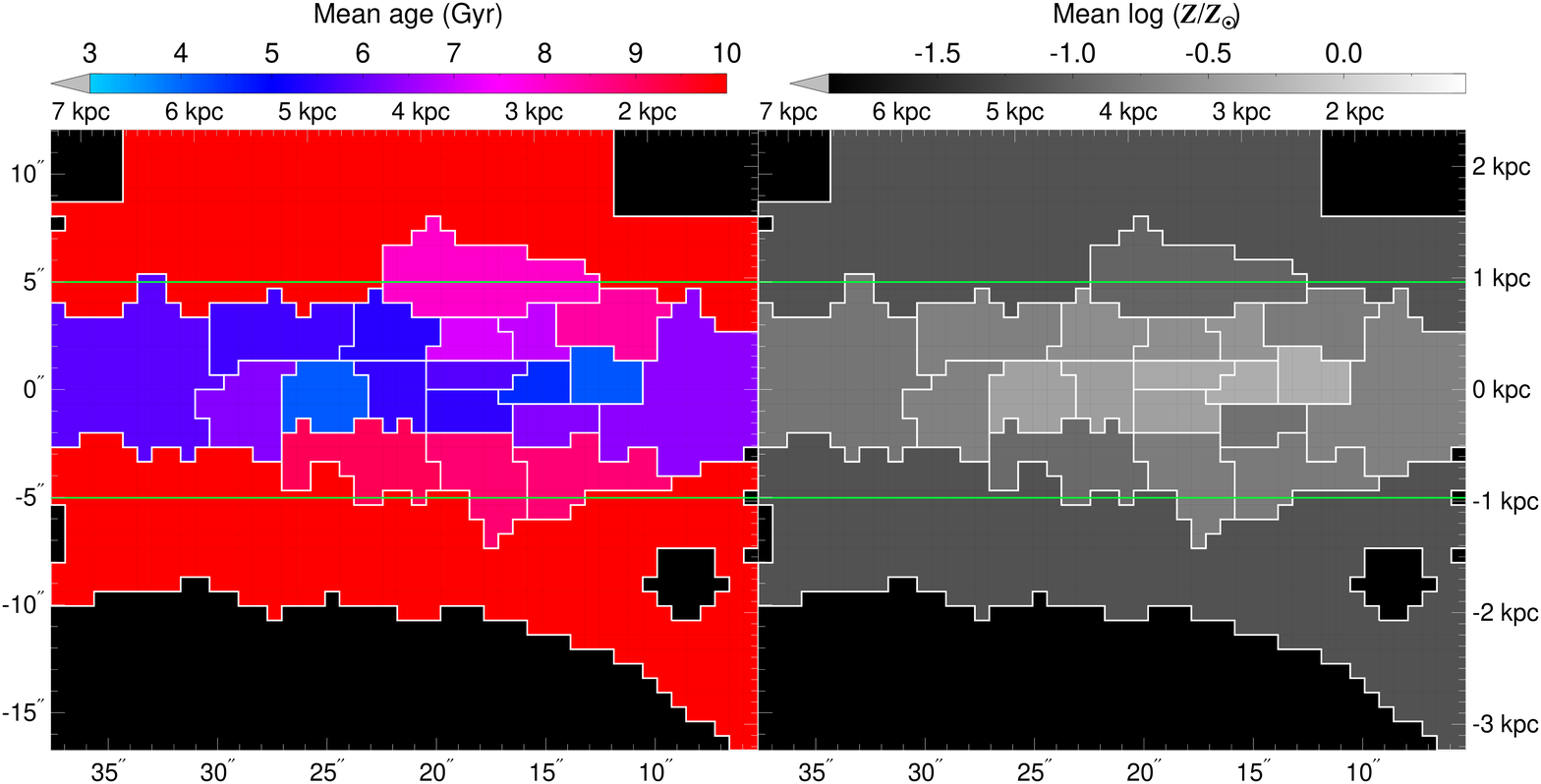}
  \end{center}
\caption{\label{means} Mean stellar age (left) and metallicity (right) maps. The white lines indicate the tesselation. The horizontal solid green lines indicate the height above which 90\% of the light comes from the thick disc.}  
\end{figure*}

To study the stellar populations, we manually created new spacial bins by combining those used in the velocity map. Before summing all the spectra from the low-$S/N$ bins to create the high-$S/N$ binning, we shifted their wavelengths so they all corresponded to a $2500\,{\rm km\,s^{-1}}$ recession velocity.  We made the bins roughly rectangular, with the long side in the direction of the plane of the galaxy, so different heights could be studied. To achieve the required $S/N$ for the thick disc, we summed the light from the two sides of the mid-plane. Manual binning is essential because the Voronoi tesselation tries to make ``compact'' or ``round'' bins \citep{CAP03}. This second binning yielded spectra with a $S/N\sim80$ in the $\lambda=4500\,\AA$ to $\lambda=5000\,\AA$ spectral range.

To obtain the stellar population map, the spectra were again fitted using pPXF. We used a grid of MILES stellar population templates ranging from 63\,Myr to 17.8\,Gyr (50 age bins) and from ${\rm log}\,\left(Z/Z_{\bigodot}\right)=-1.71$ to ${\rm log}\,\left(Z/Z_{\bigodot}\right)=0.22$ (six metallicity bins). Thus, the total template number was 300. A regularization factor of REGUL=200 was used to smooth the stellar populations. The continuum was corrected with multiplicative Legendre polynomia (for details on the regularization and the use of Legendre corrective polynomia, see the pPXF documentation). These corrections are essential because of the dust reddening close to the mid-plane and because the blue side of the spectrum has little signal, so it is prone to photometric calibration errors. We tested which polynomial order would work best for our observations. The thick disc fit, where dust reddening is likely to be negligible, is very stable with respect to changes of the order of the polynomial (which indicates little dust affecting the spectra and also a good data calibration that makes correcting polynomia unnecessary). The fitted thin disc stellar populations are very dependent on the order for orders smaller than $\sim10$. For larger orders the fitted stellar populations change little with changes in order, which probably indicates that the continuum shape is properly corrected for differential dust absorption and no further orders are required. The results presented here are fitted with an order 14 polynomial, which is comparable to what others have used with VIMOS data \citep[for example][used 25]{CHI09}. The stellar population map is presented in Fig.~\ref{populations}. The fit to the spectra in two of our tiles are shown as an example in Fig.~\ref{examplespec}. In Fig.~\ref{means}, mean age and metallicity maps are shown.

\section{Interpretation of the data}

\subsection{Kinematics}
\label{kinem}

\begin{figure}
\begin{center}
  \includegraphics[width=0.48\textwidth]{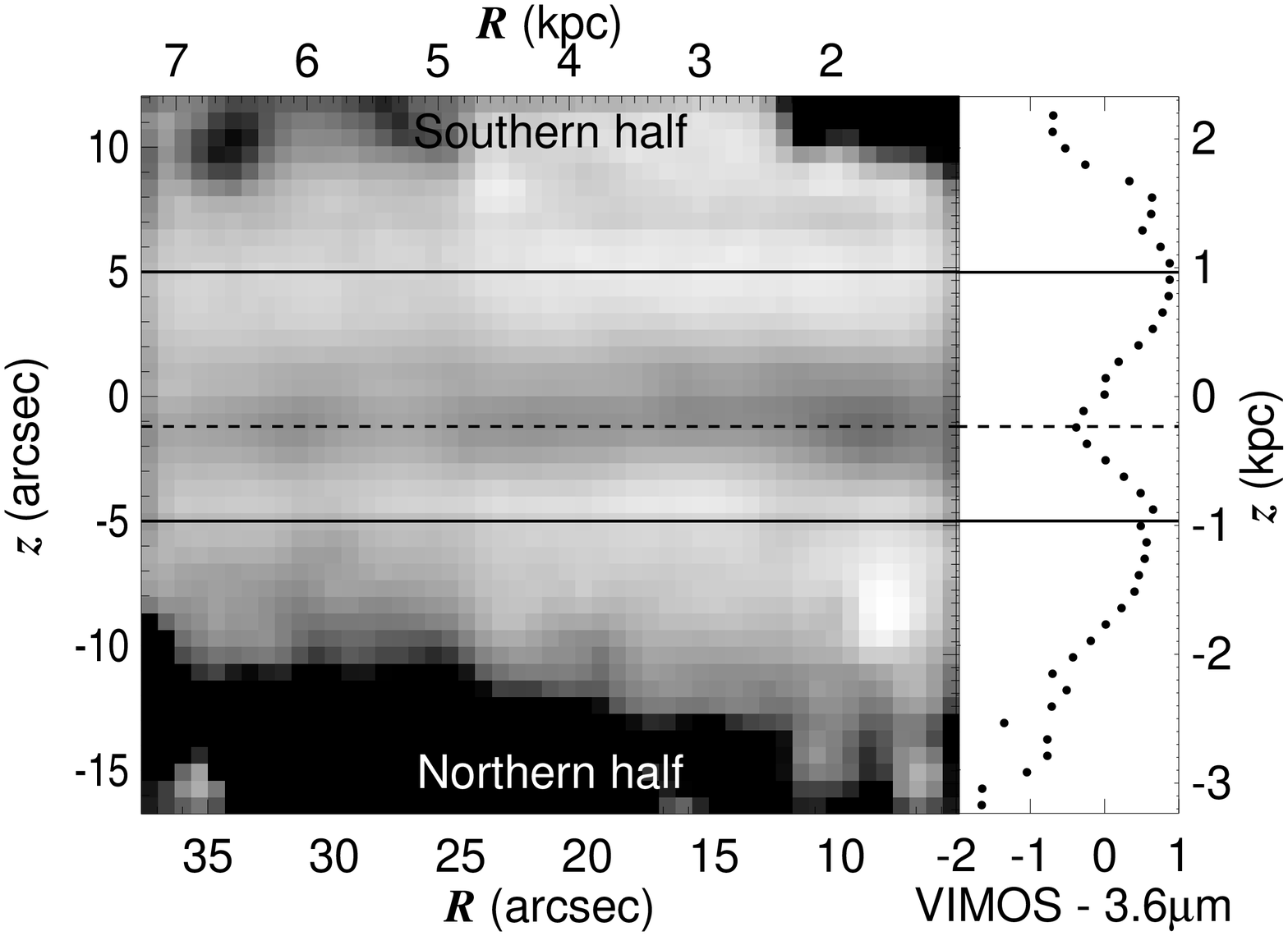}
  \end{center}
  \caption{\label{colour} Left: VIMOS$-3.6\,\mu{\rm m}$ colour map. The VIMOS image was degraded to the S$^4$G resolution. Right: median VIMOS$-3.6\,\mu{\rm m}$ as a function of the height with an arbitrary zero point. The solid horizontal lines indicate the height above which $\sim90\%$ of the light comes from the thick disc. The dashed horizontal line indicates the reddest height within the thin disc dominated region.} 
\end{figure}

\begin{figure}
\begin{center}
  \includegraphics[width=0.48\textwidth]{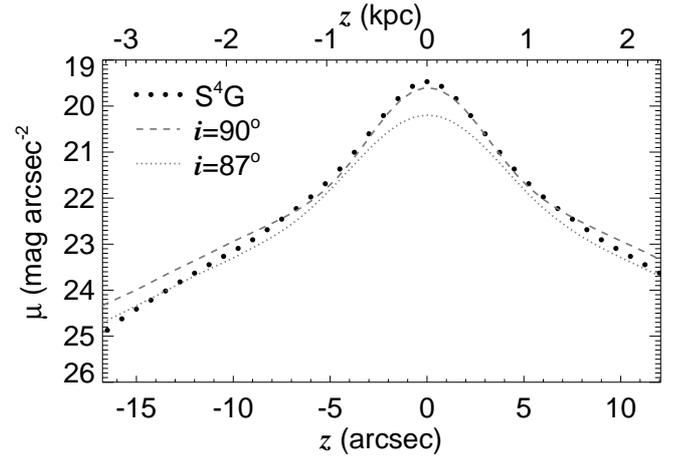}
  \end{center}
  \caption{\label{match} Luminosity profiles of the observed galaxy (large black dots) and the model galaxies (grey lines) in the $0.08\,r_{25}-0.69\,r_{25}$ axial range (dust extinction set to zero). Two models, one with $i=90^{\rm o}$ (dash), and one with $i=87^{\rm o}$ (dots), are shown. The zero point of the model profiles is arbitrary.}  
\end{figure}

\begin{figure*}
\begin{center}
  \includegraphics[width=0.95\textwidth]{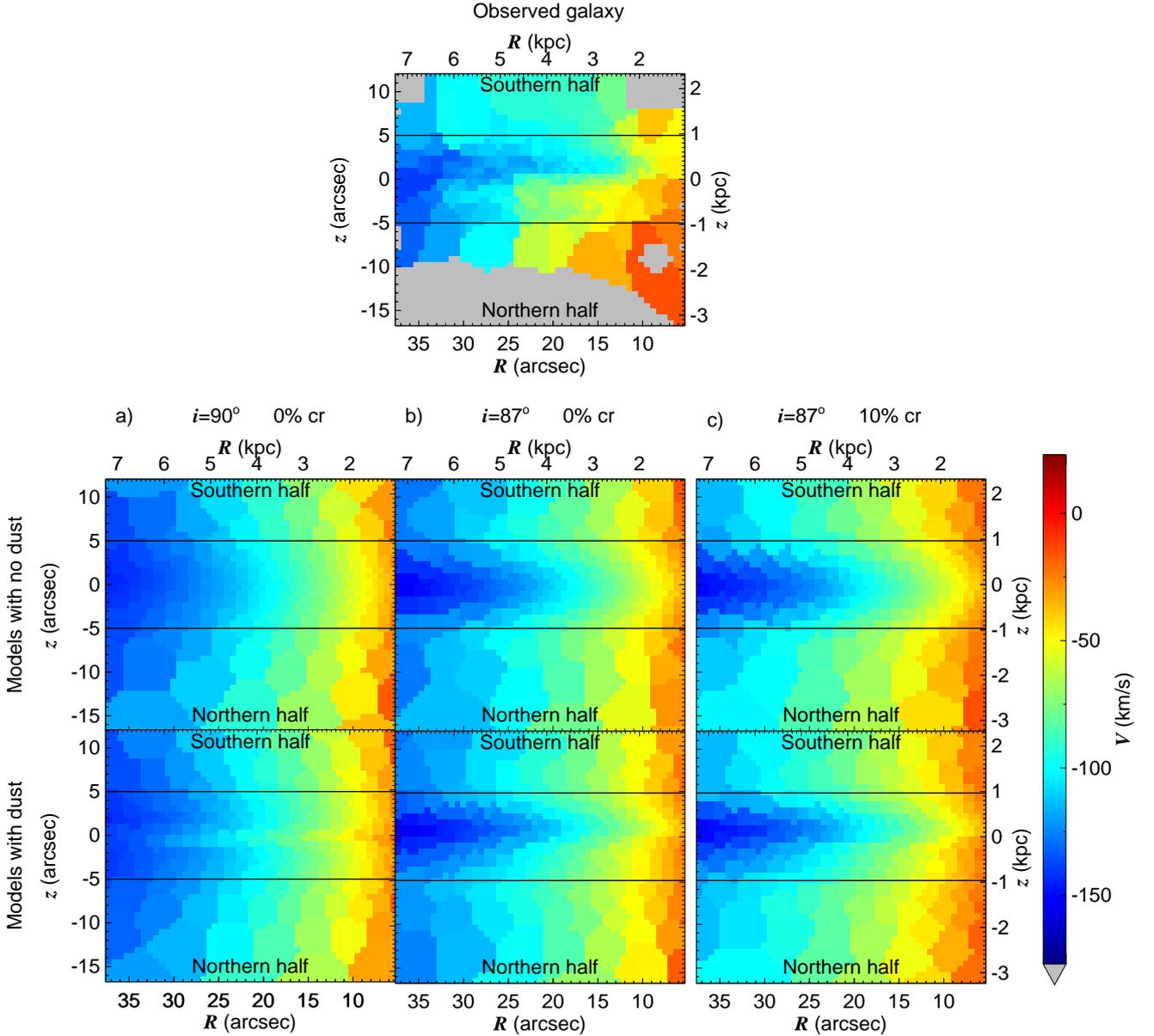}
  \end{center}
  \caption{\label{modelkinem} Top: observed velocity map of ESO~533-4. Middle and bottom rows: set of velocity maps from models designed to mimic the region shown in Fig.~\ref{velocity}. The three columns correspond to: a) a perfectly edge-on galaxy with no counter-rotating material; b) a galaxy $3^{\rm o}$ away from edge-on with no counter-rotating material; and c) a galaxy $3^{\rm o}$ away from edge-on with 10\% of counter-rotating material in the thick disc. The top row corresponds to models with no dust and the bottom row corresponds to models with some amount of dust (see text). The horizontal lines indicate the height above which $\sim90\%$ of the light comes from the thick disc.}  
\end{figure*}

\begin{figure}
\begin{center}
  \includegraphics[width=0.48\textwidth]{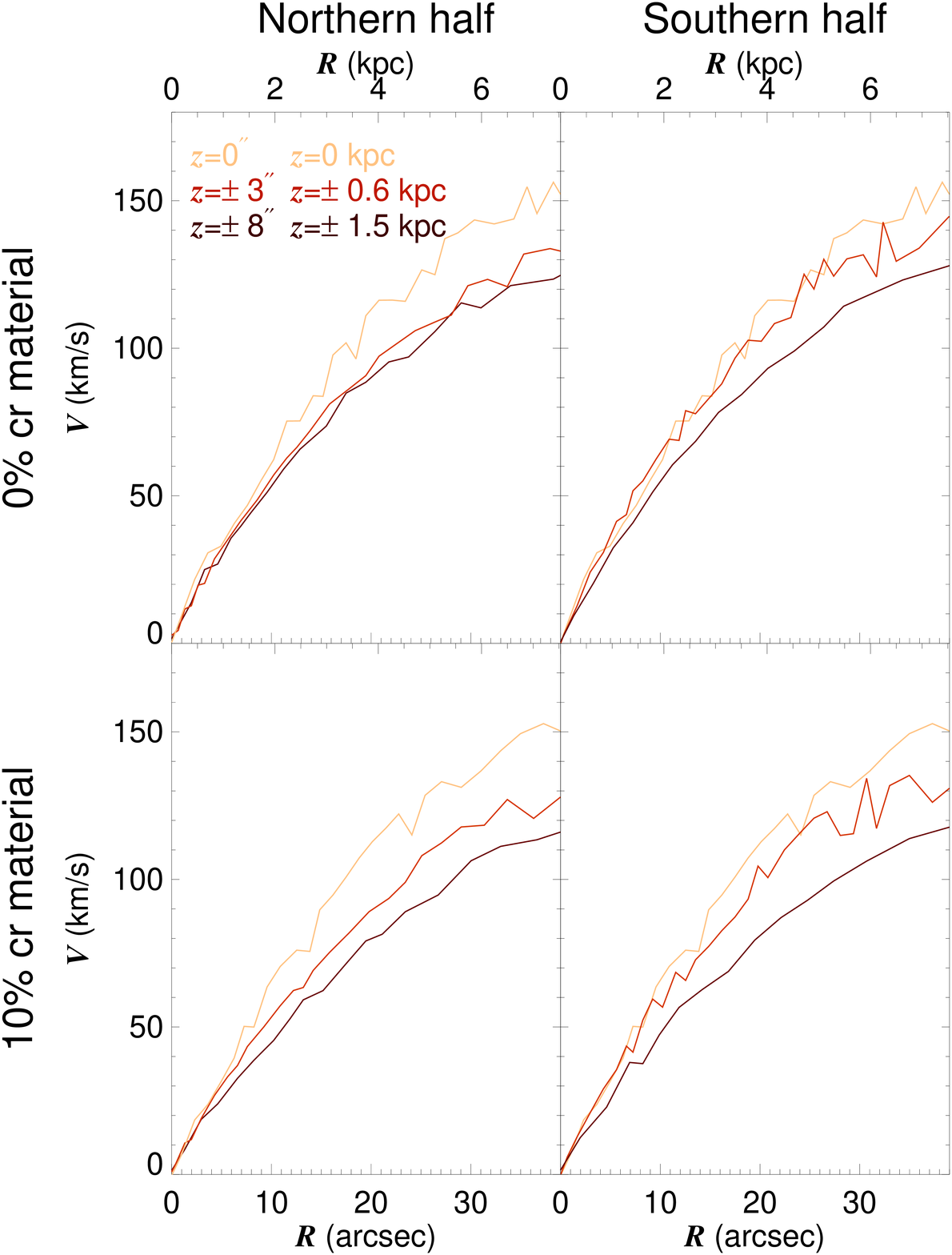}
  \end{center}
  \caption{\label{modelprofile} Rotation curves of models at different heights as indicated by the colours. The northern half of the disc corresponds to the bottom half of the disc in Fig.~\ref{modelkinem}. The top row rotation curves correspond to a model with $i=87^{\rm o}$, dust and 0\% of counter-rotating material, and the bottom row corresponds to similar model  with 10\% of counter-rotating material in the thick disc.} 
\end{figure}

We find an asymmetry in the velocity map: within the thin disc at a given radius, the southern (upper) half of the galaxy seems to rotate faster than the northern (lower) half (Fig.~\ref{velocity} and the $z=\pm3^{\prime\prime}$ or $z=\pm0.6$\,kpc rotation curves in Fig.~\ref{velprof}). Possible reasons for the asymmetry are the presence of a dust lane in a not perfectly edge-on disc and features that are not symmetric with respect to the mid-plane \citep[e.g. some phases of a buckling bar such as those seen in][]{MAR06}. The latter possibility, however, is unlikely because there is no sign of a strong bar in ESO~533-4 \citep[boxy/peanut inner isophotes;][]{COM81}.

The rotation curves for the northern half of the galaxy (below the mid-plane in Fig.~\ref{velocity}) decrease monotonically in amplitude as $z$ increases. For the southern half, this in only true for $R\gtrsim25^{\prime\prime}$ (a striking feature of the rotation curves is that for the southern half of the galaxy, we find the velocities at $z=3^{\prime\prime}$ to be larger than those at $z=0$ in the $12^{\prime\prime}-25^{\prime\prime}$ or $2.3-4.8$\,kpc axial extent). The difference in velocities between the mid-plane and $|z|=8^{\prime\prime}$ ($z=\pm1.5$\,kpc) is $10-30\,{\rm km\,s^{-1}}$ at the radius where the maximum circular velocity is reached. The apparent rotational lag of the thick disc must at least partly be a consequence of a stronger asymmetric drift caused by the larger velocity dispersion of the thick disc. Part of the lag might be caused by the presence of counter-rotating stars. Finally, part of the apparent lag might be because we may be looking at a disc that is not perfectly edge-on.

To check for the causes of the asymmetries in the velocity map and for those of the lag of the thick disc, we ran a set of models using GADGET-2 \citep{SPRIN05}. The initial models were created with the GalactICS software \citep{KUIJ95}. They included a thin disc and a thick disc, which had similar scale lengths. The scale height of the thick disc was $\sim4.9$ times larger than that of the thin disc and the mass ratio was $1.1/1.5$ in agreement with the values in Table~\ref{summary} and the fits in Fig.~\ref{thinthick}. Each of the discs was composed of $1.03\times10^6$\,particles with a Toomre parameter $Q=1.3$. We also included a small spherical CMC, with a mass 50 times smaller than that of the sum of the two discs,  formed of $2\times10^4$\,particles. Finally, we embedded the whole system into a dark matter halo made of $1.08\times10^6$\,particles. The distribution of dark matter particles was set to make a roughly flat rotation curve at large radii. At a radius of two scale lengths the disc contributed to about 40\% of the total radial force. The baryonic mass of our model scaled to the observed axial and vertical (Figure~\ref{match}) profiles and the rotational velocity amplitude is $2.9\times10^{10}\,M_{\bigodot}$. This is close to the actual mass of ESO~533-4. Models with different dark matter fractions could be created, but then having a good simultaneous vertical and axial match between the models and observations would not have been possible. The model snapshots shown here are for models that have been run for over ten rotations at all the radii covered by the VIMOS observations. This guarantees that the thin and  thick discs have settled to equilibrium.

To account for the effect of dust, we assumed it would follow the distribution of thin disc stars except for the fact that we considered a scale height that was three times smaller. We arrived at the ratio of three times smaller  because we assumed dust to be mixed with dense gas. In the Milky Way, dense gas has a scale height of $\sim100$\,pc \citep{LAN14} whereas the thin disc scale height is $\sim300$\,pc \citep{GIL83}. We assumed the dust would cause a grey absorption, and we studied several dust densities. The results shown here correspond to an optical depth of $\tau=8$ to the centre of the galaxy when looking at it from a perfectly edge-on view.

We checked whether ESO~533-4 has some measurable deviation from an edge-on orientation by looking at its colour map (Fig.~\ref{colour}). We found the redder sections of the thin disc to lie slightly below the mid-plane ($\sim1^{\prime\prime}$). This implies that the near side of the galaxy is on top, which corresponds to the southern side of the observed galaxy in Fig.~\ref{velocity}. We find that a $\sim1^{\prime\prime}$ deviation of the dust lane corresponds to a galaxy inclination $i\approx85^{\rm o}$ in our models. We simulated colour maps by comparing the luminosity profiles of models with and without dust. ESO~533-4's inclination, however, is likely to be closer to $90^{\rm o}$ given the extremely thin aspect of the disc and the lack of evidence for spiral features (Fig.~\ref{ESO533-4}). If the dust distribution were less centrally concentrated than in our models, deviations of only $2-3^{\rm o}$ away from edge-on would be enough to explain the vertical shift of the dust lane. There are clear examples of galaxies where the dust distribution is more extended than that of the stellar disc \citep[e.g. the Sombrero Galaxy or the system studied in][]{HOL09}.

To create synthetic velocity maps, we assigned the MILES spectrum of a 2.5\,Gyr star with solar metallicity to thin disc particles. For the thick disc, we used the spectrum of a 10\,Gyr star with metallicity $Z=0.2\,Z_{\bigodot}$. We created synthetic images of the galaxy by projecting them and convolving them with a Gaussian PSF with a FWHM of $1\farcs2$. We then used the Voronoi tesselation software to create a tesselation matching that obtained for the VIMOS data (i.e. the density of tiles in the thin disc is similar). We then obtained the velocity maps exactly as we did for the VIMOS data.

Figure~\ref{modelkinem} shows the velocity maps for several models in an area approximately matching that of the VIMOS observations. The top row in the figure corresponds to galaxies with no dust and the bottom row corresponds to galaxies with dust. Models with no dust are symmetric with respect to the mid-plane. Models with dust, when not seen exactly edge-on, have some degree of asymmetry.

The left column in Fig.~\ref{modelkinem} corresponds to models with a perfectly edge-on view of the galaxy. The optical depth is the largest close to the mid-plane. Dust attenuates the light from the point which is closer to the galactic centre (that with the largest velocity along the line of sight) more than that of stars closer to the observer. Thus, the dust attenuation causes an apparent slowing of the stars around $z=0$.

The middle and the right columns in Fig.~\ref{modelkinem} show that dust is able to create asymmetries when looking at galaxies that are not perfectly edge-on (inclination $i=87^{\rm o}$ in the example shown here). The larger attenuation on one side of the galaxy causes the asymmetry in the velocity maps: the maximum velocity at a given radius is not found at the mid-plane, but slightly above it. The deviations from symmetry in our models with $i=87^{\rm o}$ and dust are in qualitative agreement with the observations (Figure~\ref{velocity}).

To find the cause of the apparent lag in the thick disc of ESO~533-4 (asymmetric drift, galaxy inclination, and/or presence of counter-rotating stars), we created rotation curves at different heights for the models with the same approach we used for the VIMOS data (Figure~\ref{modelprofile}). We find that at the radius where $v(z=0)\approx150\,{\rm km\,s^{-1}}$, the difference in velocity between the mid-plane and $z=\pm8^{\prime\prime}$ ($z=\pm1.5$\,kpc) is $\sim20\,{\rm km\,s^{-1}}$ for a model with no counter-rotating material and $\sim30\,{\rm km\,s^{-1}}$ for a galaxy whose thick disc has  10\% of counter-rotating material.  These numbers are similar to those observed in ESO~533-4 and indicate that the thick disc has little or no counter-rotating material. The effect of the asymmetric drift plus the disc orientation (slightly away from edge-on) are enough to explain the thick disc lag.

In Fig.~\ref{modelprofile} we see how, in the southern half, the $z=3^{\prime\prime}$ ($z=\pm0.6$\,kpc) curve overlaps with that at $z=0$ for some radial extent. This is because dust absorption is larger at $z=0$ than at $z=3^{\prime\prime}$. By increasing dust absorption it is possible to obtain the $z=3^{\prime\prime}$ curve above the $z=0$ curve for some axial extent just as in the observations. This effect is therefore most likely due to dust. 

Our models are not able to reproduce  the rotation curves of the northern half of ESO~533-4 for $R<25^{\prime\prime}$ ($R<4.8$\,kpc), where the thick disc lags by $30\,{\rm km\,s^{-1}}$ or more with respect to the mid-plane. The mismatch between the thick disc and the $z=0$ rotation curves could be due to some fraction of counter-rotating stars in the inner parts of the galaxy. However, if that were the case it would be difficult to explain why the counter-rotating material shows up only on one side of the galaxy.

\subsection{Stellar populations}

\label{pops}

\begin{figure}
\begin{center}
  \includegraphics[width=0.48\textwidth]{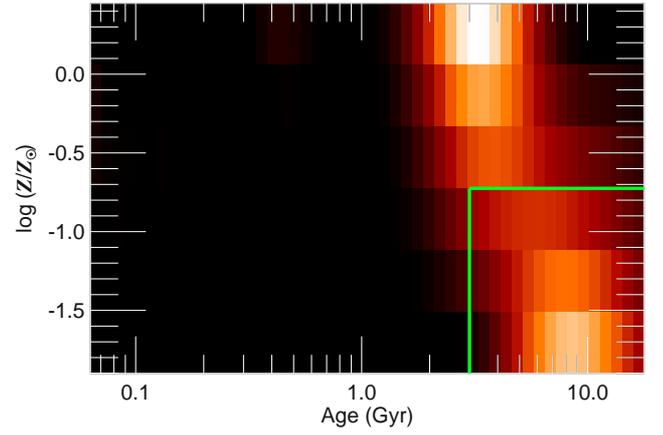}
  \end{center}
  \caption{\label{example} One of the subplots in Fig.~\ref{populations} (the rightmost). The green lines indicate the divisions between the thick disc population (bottom right) and the thin disc (rest of the plot).} 
\end{figure}

\begin{figure}
\begin{center}
  \includegraphics[width=0.48\textwidth]{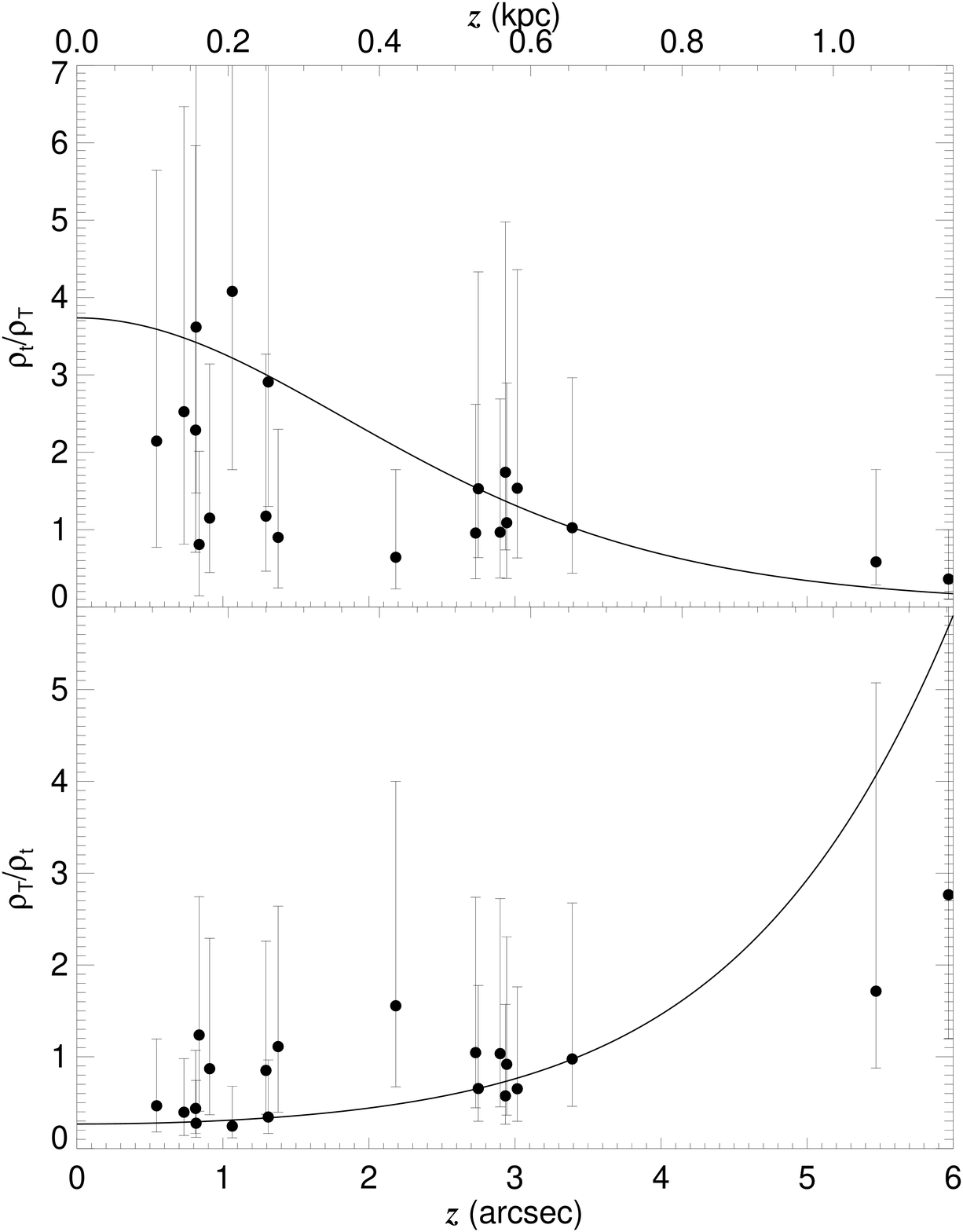}
  \end{center}
  \caption{\label{ratio} Ratio between the masses of the thin and thick disc stellar populations (top panel) and its inverse (bottom panel), as a function of the height $|z|$ of the Fig.~\ref{populations} tile. The lines indicate the same ratios according to the fit to the light profiles (Figure~\ref{velocity}).} 
\end{figure}

We  distinguish two distinct stellar populations in the subplots in Fig.~\ref{populations}. We find:
\begin{itemize}
 \item an old ($\sim10\,{\rm Gyr}$) metal-poor population and
 \item a relatively young metal-rich population.
\end{itemize}
Also, traces of young metal-poor populations, either a fit artefact or associated with the outskirts of the thin disc, are seen in a few of the tiles (this is most evident in the tiles 2, 4, and 6).

To quantify the fraction of mass associated with each of these stellar populations in each tile, we divided the diagrams in two regions, as seen in Fig.~\ref{example}. The limit between the thin and the thick discs is set at ${\rm log}\,(Z/Z_{\bigodot})=-0.7$ for ages larger than 3\,Gyr. All populations younger than 3\,Gyr are considered to belong to the thin disc. While a $t=3\,{\rm Gyr}$ stellar population is young compared to what one would expect for a thick disc population, the regularization introduced by pPXF causes stellar populations with peak ages at $t\sim10\,{\rm Gyr}$ to extend down to smaller ages.

The first of the populations in the list is naturally associated with the thick disc, and the second is likely to be related to the thin disc. To further confirm this, we find that the old metal-poor population  dominates the mass budget in the bin with the largest height, whereas the younger metal-rich population dominates close to the mid-plane.

In many tiles, the stellar population fits show that the thin and  thick disc populations have two distinct maxima in the ${\rm Age}-{\rm log}\,(Z/Z_{\bigodot})$ plane. In these tiles, the thin and thick disc stars seem to belong to two distinct populations. In other tiles (particularly 4, 6, 11, and 12 in Fig.~\ref{populations}) the two populations are not so clearly separated. The choice of the regularization parameter introduces a certain arbitrarity in the interpretation of the age-metallicity plane. However, this is a necessary evil, since a certain amount of regularization is needed to smooth the otherwise even more arbitrary solutions of a difficult inverse problem (see pPXF online documentation). While varying the regularization parameter between REGUL=100 and REGUL=250 does not drastically change the results, higher values can wash out significantly the peaks. The chosen regularization value of REGUL=200 is similar to the value REGUL=250 used in the only source where, to our knowledge, a concrete value of this parameter has been explicitly stated (the example codes from the pPXF website).

If we assume that the stars are divided into a thin and  thick disc population (as indicated by many of the ${\rm Age}-{\rm log}\,(Z/Z_{\bigodot})$ plots in Fig.~\ref{populations}), we can test the assumptions made when fitting the luminosity profiles to decompose them into the thin and thick disc components, as  in \citet{CO11B, CO11C, CO12, CO14}. We calculated for each tile the mass that corresponds to the thin and  thick discs according to the fit in the left panel of Fig.~\ref{velocity}. To accomplish this, we weighted the spaxels accounting both for the vertical and the axial luminosity profiles \citep[axial luminosity profile from][]{CO12}. Because of the broadness of the peaks that correspond to the stellar populations of the thin and the thick discs (see Fig.~\ref{example}), the thick disc stellar population might extend to larger metallicities that the ones considered here (limit set at ${\rm log}\,(Z/Z_{\bigodot})=-0.7$). Likewise, the thin disc stellar population could extend to smaller metallicties than the ones considered here. To account for this, we considered the possibilities that the thick disc population extended up to ${\rm log}\,(Z/Z_{\bigodot})=-0.3$ or only up to ${\rm log}\,(Z/Z_{\bigodot})=-1.1$ (for ages larger than 3\,Gyr). We also considered the possibilities that the thin disc extended either down to ${\rm log}\,(Z/Z_{\bigodot})=-1.1$ or only down to ${\rm log}\,(Z/Z_{\bigodot})=-0.3$ for ages larger than 3\,Gyr. This is what has been used to calculate the error bars in Fig.~\ref{ratio}. The figure shows that the agreement between the results obtained from the stellar population maps and that from our thin/thick disc modelling is good, especially if we take the uncertainties and assumptions made in the modelling process into account.

\section{Summary of the results, discussion, and conclusions}

In this paper we have made an IFU study of the thin and the thick discs of ESO~533-4 with VIMOS. The main findings are:
\begin{itemize}
 \item The velocity field of ESO~533-4 is compatible with that of a galaxy inclined a few degrees from edge-on with a significant dust absorption in the equatorial plane. The near side of the galaxy is the southern one.
 \item The apparent velocity lag of the thick disc can be explained by asymmetric drift and because of the projection effects arising from studying a disc a few degrees ($2-3^{\rm o}$) away from edge-on. Few or no counter-rotating stars are needed to account for the asymmetric drift of the thick disc.
 \item The thick disc is made of old ($\sim10\,{\rm Gyr}$) and metal-poor stars. The thin disc is made of younger metal-rich stars. This is manifested as a distribution with two distinct peaks in the ${\rm Age}-{\rm log}\,(Z/Z_{\bigodot})$ plane for the majority of the regions of ESO~533-4 we studied. This might indicate that the stellar populations of the thin and thick discs are distinct. In several regions (at least 4 out of 20) there is no distinct peaks, but instead a continuum distribution of stellar populations from old metal-poor stars to younger metal-rich stars.
 \item The relative fractions of mass in the thin and the thick discs at different heights, according to our stellar population analysis, agree with those obtained from our thin/thick disc fits of mid-infrared luminosity profiles.
\end{itemize}

We first examine the lack of a significant fraction of counter-rotating stars in the thick disc. If the thick disc stars were accreted in a succession of minor mergers, we would expect similar fractions of prograde and retrograde stars, unlike what we observe. If the thick discs were created in a few relatively major mergers \citep[as in e.g.][]{READ08} or as a result of the fusion of a few protogalactic mergers, as in \citet{BROOK04}, there would be no guarantee that a particular galaxy would have counter-rotating stars. Indeed, in the extreme case where only one such an event occurs, there is a 50\% chance of having a prograde merger and a 50\% chance of having a retrograde merger. Thus, the lack of counter-rotating material in ESO-533-4 would not rule out a merger origin. If we consider the 1D spectroscopy study by \citet{YOA08A} and our knowledge of the Milky Way kinematics \citep{CHI00}, we find that none of the five massive galaxies ($v_{\rm c}\gtrsim120\,{\rm km\,s^{-1}}$) for which thick disc kinematics are known has a counter-rotating disc. Therefore, if all thick discs in high-mass galaxies share a similar history \citep[which is a reasonable assumption, given that the ratio of the masses of the thin and thick discs are roughly similar for most of the high-mass galaxies;][]{CO14}, a major merger origin is disfavoured by the currently available data. We thus tentatively propose that the thick disc of ESO~533-4 has formed either from dynamical heating or that it was born dynamically hot in a turbulent disc with a large star formation rate.

To distinguish between those two mechanisms, we can use our stellar population data. We see that in most of the tiles in Fig.~\ref{populations}, the thin and the thick disc populations are clearly differentiated in the ${\rm Age}-{\rm log}\,(Z/Z_{\bigodot})$ plane, which would go against a secular heating scenario where a continuous transition between the thin and the thick discs is expected. Instead, it would favour a scenario in which the thick disc became thick at high redshift in a relatively fast process. We however caution against over-interpreting the stellar population results because of the extremely dusty conditions under which the fitting is done and because of their dependence of the smoothing introduced by the regularization of the data. Confirmation of our stellar population results with longer wavelength spectra is thus desirable.

Another and probably more robust way to distinguish between the two mechanisms compatible with our kinematical data is hinted at in \citet{MIN15}. In their model, age gradients naturally appear in the thick disc-dominated region. These gradients would immediately confirm or infirm the formation scenario where discs are made at high redshift. Unfortunately, because of $S/N$ constrains, we can only make a single bin out of  the entire thick disc dominated region, which prevents us from testing for the presence of an age gradient. Even if a larger number of thick disc bins could be made, the effect predicted by \citet{MIN15} would be very subtle because our observations extend only out to about 1.5 disc scale lengths \citep{CO12}, whereas a large age difference (of the order of $2-4$\,Gyr) would in theory be observed only over ranges of $\sim3$ scale lengths. Larger field of view observations (e.g. with MUSE) are thus needed to  settle this issue definitely.

The conclusion is that, if we assume that all thick discs in high-mass galaxies share a common origin, observations favour a dynamical heating or turbulent disc origin for the thick disc in ESO~533-4. A minor merger origin can be discarded, whereas a major merger origin is possible if a variety of scenarios are able to produce a thick disc in high-mass galaxies. The stellar population map of ESO~533-4 is affected by dust absorption and should be interpreted with caution. Recovering the stellar populations from a spectrum is an ill-posed problem which requires the choice of a regularization parameter to smooth the solution. Having said that, the stellar population fits seem to indicate that this particular galaxy has differentiated thin and thick disc stellar populations, as expected from a short event (either in an early turbulent disc or in a major merger). It thus seems to disfavour a secular evolution origin. A study of the axial gradients of the stellar ages in the thick disc is required to confirm this last assertion.

\begin{acknowledgements}

We thank our referee for helping to improve the interpretation of the data. We thank Dr.~Michele Cappellari for his help  when fitting stellar populations. We thank Dr.~Valentin Ivanov for his help during a very complicated observing stint. We thank Dr.~Julio Carballo-Bello for pointing out at some useful references. We thank Dr.~Fr\'ed\'eric Bournaud for providing details about his simulations. This work is based on observations and archival data made with the Spitzer Space Telescope, which is operated by the Jet Propulsion Laboratory, California Institute of Technology under a contract with NASA. We are grateful to the dedicated staff at the Spitzer Science Center for their help and support in planning and execution of this Exploration Science program. We also gratefully acknowledge support from NASA JPL/Spitzer grant RSA 1374189 provided for the S$^4$G project. This research has made use of SAOImage DS9, developed by Smithsonian Astrophysical Observatory. This research has made use of the NASA/IPAC Extragalactic Database (NED), which is operated by the Jet Propulsion Laboratory, California Institute of Technology, under contract with the National Aeronautics and Space Administration. We acknowledge the usage of the HyperLeda database (http://leda.univ-lyon1.fr). SC, HS, and EL acknowledge support from the Academy of Finland. JJ thanks the ARC for financial support via DP130100388. the authors acknowledge support for the FP7 Marie Curie Actions of the European Commission, via the Initial Training Network DAGAL under REA grant agreement number 289313.

\end{acknowledgements}

\bibliographystyle{aa}
\bibliography{ESO533-004}

\end{document}